\documentclass[11pt]{article}
\usepackage[utf8]{inputenc}
\usepackage[T1]{fontenc}
\usepackage{graphicx}
\usepackage{grffile}
\usepackage{longtable}
\usepackage{wrapfig}
\usepackage{rotating}
\usepackage[normalem]{ulem}
\usepackage{amsmath}
\usepackage{textcomp}
\usepackage{amssymb}
\usepackage{capt-of}
\usepackage{hyperref}
\graphicspath{{figures/}}
\usepackage{stmaryrd}
\usepackage{bbding}

\let\saveAsterisk\Asterisk
\let\Asterisk\relax
\usepackage{mathabx}
\let\Asterisk\saveAsterisk

\author{Ryan Bernstein}
\date{\today}
\title{Static Analysis for Probabilistic Programs}
\hypersetup{
 pdfauthor={Ryan Bernstein},
 pdftitle={Static Analysis for Probabilistic Programs},
 pdfkeywords={},
 pdfsubject={},
 pdfcreator={Emacs 25.3.1 (Org mode 9.1.13)}, 
 pdflang={English}}
\begin{document}

\maketitle

\begin{abstract}
Probabilistic programming is a powerful abstraction for statistical machine learning.  Applying static analysis methods to probabilistic programs could serve to optimize the learning process, automatically verify properties of models, and improve the programming interface for users.  This field of static analysis for probabilistic programming (SAPP) is young and unorganized, consisting of a constellation of techniques with various goals and limitations.  The primary aim of this work is to synthesize the major contributions of the SAPP field within an organizing structure and context.  We provide technical background for static analysis and probabilistic programming, suggest a functional taxonomy for probabilistic programming languages, and analyze the applicability of major ideas in the SAPP field.  We conclude that, while current static analysis techniques for probabilistic programs have practical limitations, there are a number of future directions with high potential to improve the state of statistical machine learning.
\end{abstract}

\tableofcontents

\section{Introduction}
\label{sec:org88e9099}

The idea of statically analyzing the source code of probabilistic programs is relatively new, because probabilistic programming is relatively new. The idea has cropped up in the form of a disparate set of approaches that have been presented with varying purpose, technique and context of application. Thus far, there has been no systematic study of this field of static analysis of probabilistic programs as a whole. This paper will attempt to fill that role, with the following goals in mind:
\begin{itemize}
\item To give a context and motivation for the application of static analysis to probabilistic programs, such that the reader may understand and appreciate the important ideas in the field.
\item To suggest some organization for the space of the field, as a means to locate ideas within the field.
\item To present, categorize and discuss some existing ideas that have been explored in this space.
\item To identify places where existing work could be fruitfully extended, and to suggest directions of potentially high impact for the future.
\end{itemize}
This paper attempts to cover a representative set of probabilistic programming languages, but uses the popular Stan language \cite{Carpenter2017StanAP} as the primary reference point. 
\subsection{Introduction to Probabilistic Programming}
\label{sec:org1b325ef}

Probabilistic programming provides a powerful means for specifying and computing with probabilistic models.

A \textit{probabilistic model} is a probability distribution over some known variables, called \textit{data} or \textit{observations}, and unknown variables, called \textit{model parameters} (or just \textit{parameters}). Probabilistic models underly statistical analyses and allow the analyst to generalize conclusions about quantities of interest beyond the observed data.

Often, the goal of writing a probabilistic model is to find the \textit{posterior} probability distribution, also called the posterior distribution. The posterior distribution is the probability distribution over the parameters conditioned on the data, representing our updated beliefs about the model parameters given the new information found in the data.

A probabilistic program expresses a probabilistic model by defining a \textit{joint} distribution. A joint distribution maps each possible pair of intantiations of data and parameters to a positive real number which corresponds to how "likely" that pair is to occur according to the model. This mapping is called the \textit{density} function of the distribution\footnote{Strictly speaking, the density function is a ratio between a distribution and a base measure, and some distributions do not have densities. Except when we deal explicitly in terms of measure theory, we will assume that every distribution has a density with respect to the Lebesgue measure. In that case, it is reasonable to think of the density as being a real number proportional to the probability.}. When a distribution's density sums (or integrates) to one, it is called a \textit{normalized} distribution or a probability distribution. The joint distribution represented by a probabilistic program does not need to be normalized.

It is very convenient and flexible to write a probabilistic model as an unnormalized joint distribution. One can:
\begin{itemize}
\item Describe a process for generating the data in terms of the parameters, along with a distribution representing one's prior beliefs about the parameters. This process corresponds to constructing a joint distribution by multiplying a likelihood and a prior.
\item Simply specify a joint distribution directly as a function of the data and model parameters.
\end{itemize}

When the data are plugged into the (potentially unnormalized) joint distribution, we get an unnormalized distribution which is proportional to the posterior probability distribution. This is a corollary of Bayes' rule.

This task of recovering the posterior probability distribution from a proportional unnormalized distribution is called \textit{posterior inference}, and it is the core computational challenge of probabilistic programming (and much of computational Bayesian statistics).
Unfortunately, exact posterior inference is computationally intractable for most real-world problems, since it involves integrating over the whole space of parameters.
We usually rely instead on approximate posterior inference methods. Some common approximate methods are described in section \ref{orga2de2f1}.

The advantages of probabilistic programs are enabled by the separation of concerns between the program, which is only concerned with representing probabilistic models, and the posterior inference engine, which is only concerned with calculating the posterior. This separation has a number of advantages over typical practices in applied statistics and machine learning:
\begin{itemize}
\item Probabilistic programs are freed of details of inference, letting users operate at a high level of abstraction. This makes models easier to write, debug, and iterate on. It also enables users to write programs which clearly communicate their domain knowledge.
\item Improvements to posterior inference algorithms propagate for free to every probabilistic program, just by rerunning the model.
\item As we will see, since probabilistic programs are written in standardized, computer-readable, and semantically meaningful format, we can apply static analysis for a number of additional benefits.
\end{itemize}

\subsection{Introduction to Static Analysis}
\label{sec:org803a13d}

The term "static analysis" refers to a set of techniques for analyzing the source code of a program before it is run. The strategy is to enable as much work as possible at the point of compiling a program in order to gain some benefit every time the program is run.

Static analysis is pervasive in software engineering. For example, some common uses of static analysis are:
\begin{itemize}
\item When a C program is made faster by GCC's -O3 option, shaving seconds off the execution of a program that will be run a million times \cite{gcc}.
\item When NASA engineers prove that the Mars rover's piloting software will never divide by zero, ensuring the safety of a mission critical system \cite{NASA}.
\item When a programmer's code editor points out a mistake and suggests corrections in real time, saving hours of debugging time \cite{ide}.
\end{itemize}
We will later see that these three examples are representative of three general purposes for static analysis: optimization, verification, and usability.

\begin{figure}
	\centering
	\includegraphics[scale=0.4]{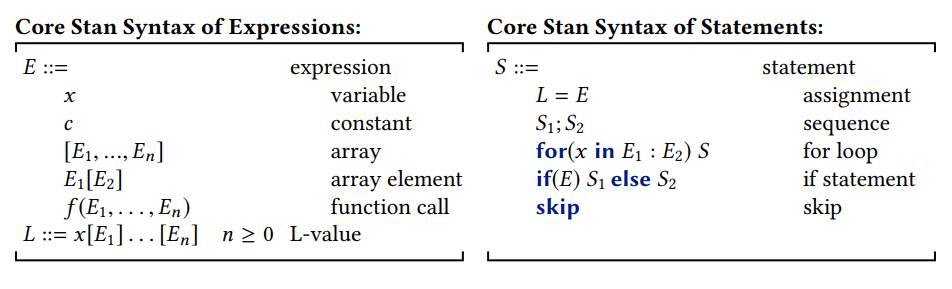}
	\caption{On the left is shown the list of grammatical rules for writing (a limited subset of) expressions in Stan. Expressions are pieces of the language which evaluate to values, such as numbers or strings. On the right is shown the list of rules for writing (a limited subset of) statements in Stan. Statements are pieces that take some imperative action, such as setting the value of a variable or running other statements in a loop. Not included here are the rules for writing a whole Stan program, which would describe the various blocks that are allowed, and how the blocks are made up of statements. The rules are read as follows: to generate a value of the type on the left hand side of the ::=, such as an E, pick one of the instances on the right hand side (separated here by lines), such as c. This means that a constant such as 1 is a valid Stan expression. Figure taken from \cite{SlicStan}.}
	\label{fig:slicstan-stan-syntax}
\end{figure}

Approaches to static analysis vary, but one unifying theme is that they work by viewing a program as a syntax tree. Syntax is the set of grammatical rules that characterizes valid programs in a language, and allows each program to be structured into a tree with the whole program as the root and individual tokens as the leaves. For example, \autoref{fig:slicstan-stan-syntax} shows a subset of the expression and statement syntax of Stan.

These motivations for static analysis are easily extended to probabilistic programming. Inferring the posterior of a model can take a lot of computing time and power, and trusting that a model is correct can be expensive and/or risky.
While typical programming can be said to have compile-time, run-time, and test-time phases, probabilistic programming can be said to have analogous compile-time, inference-time, and model checking-time phases. Each of these points could potentially benefit from static analysis using the new information available at each step.

\subsection{Purposes of applying Static Analysis to Probabilistic Programming}
\label{sec:org18a3e66}

Static analysis can be applied to probabilistic programs for a variety of reasons.

\subsubsection{Optimization}
\label{sec:org3ecd59e}

The goal of optimization is, broadly, to get more done with fewer resources.

While optimization of traditional programs is done with respect to time or memory for a single execution of a program, optimization of probabilistic programs can consider the entire process of posterior inference, perhaps with respect to the average time or memory it takes to converge on a posterior distribution. This usually includes the traditional optimization of a program as a subproblem, but also considers how the model can be analyzed or transformed to better suit it for the particular inference method in use.

A second difference from optimization of traditional programs comes from the difference in the program semantics that need to be preserved. In traditional programming, the optimized program needs to have the same behavior when given the same inputs as the original. In probabilistic programming, the program needs to represent the same posterior distribution as the original. This difference leads to additional avenues for optimizing probabilistic programs.

  The performance of probabilistic program inference is an important consideration. The performance advancements of the No-U-Turn Sampler algorithm \cite{Hoffman2014TheNS}, for example, have led to a significant number of practitioners turning to probabilistic programming for their probabilistic modeling tasks. The performance gains of variational inference, although often coming at a large cost of precision, have further expanded probabilistic programming into more applications. Fast model-to-posterior times directly enable users to complete tasks that are otherwise infeasible, and lower barriers to users expressing models they believe are most correct.
If we could perfectly optimize probabilistic programs, we could:
\begin{itemize}
\item Avoid all computation which is unnecessary, whether because it is irrelevant, overprecise, or redundant.
\item Appropriately parallelize computation for hardware such as CPUs, GPUs and TPUs.
\item Make all useful information available to the inference engine.
\end{itemize}

\subsubsection{Verification}
\label{sec:org351017d}
\label{orgac79c64}

Software verification is the task of proving properties of programs. Examples of some typical types of properties are:
\begin{itemize}
\item A variable will never be negative;
\item A loop will terminate;
\item Some behavior is insensitive to perturbation in the program's input.
\end{itemize}
Properties can be proven automatically by a variety of techniques, including by type systems or abstract interpretation, which are discussed in the technical background section.

When we extend verification to probabilistic programs, we can use the usual deterministic reasoning, but we can also extend both the properties and the proofs to include probabilistic reasoning. We can include random variables in the property, such as querying whether the expected value of a random variable will always be positive. We could also consider the probability of the truth of a property, such as asking whether a safety property holds with sufficient probability.

There are a number of types of errors in probabilistic programming that we might want to catch using verification.
\begin{itemize}
\item Depending on the probabilistic programming language, there might be any of the usual programming errors: infinite loops, null pointer references, buffer overflows and so on. This type of error can be caught by implementing the techniques of standard compilers.
\item There might be an accidental mismatch between the program and the probabilistic model the user intended to represent. Software engineers call this a logic error. This type of error might be caught by warning the user of potential common errors or by checking the program against additional assertions supplied by the user.
\item Even if a program perfectly represents the intended probabilistic model, the resulting posterior may still be incorrect if the model could not be inferred using the chosen inference method. Catching this type of error would be specific to the inference method in use.
\end{itemize}
Any progress toward catching these classes of error could automatically improve the modeling process for users en masse.

The ability in software engineering to automatically verify aspects of programs has become a vital safety feature for a variety of industries, such as healthcare, finance, and aerospace engineering. Probabilistic programming is in a unique position among machine learning and statistics tools to apply an analog of these automated proofs in data science. Given the calls for trustworthiness in machine learning and for robust statistical standards in science, any practical results in this space could have significant impact.

\subsubsection{Usability}
\label{sec:orgc8f6c5e}

Static analysis can also be used to improve the human interface to probabilistic programming. In software engineering, static analysis is used to automatically generate, transform, and give immediate feedback on programs, and makes writing good programs easier and more accessible. This ease can certainly be translated to probabilistic programming. The main difference will be that interface improvements could potentially ease the entire modeling, inference and model criticism task, of which the programming task is a subset.

Advances in usability correspond to broadening the availability of probabilistic programming to a wider audience. It could also decrease the mental overhead of all users, allowing them to spend more resources on the quality of their models. Usability changes could also improve the interpretability of programs, which increases their usefulness as communication devices.

\section{Technical background}
\label{sec:org1dad463}
\subsection{Posterior inference}
\label{sec:org45b650c}
\label{orga2de2f1}

Posterior inference algorithms take in a probabilistic program and observations of the known variables \footnote{Some literature assumes that the data is embedded inside the probabilistic program, and so posterior inference would not need the data to be supplied separately. This paper often treats them as being supplied separately for clarity and to agree with the usage of languages like Stan. The two options are equivalent except when the data is needed for a compile-time static analysis.}, and produce a representation of the posterior distribution over the model parameters.

There are at least two popular flavors of approximate inference: sampling-based Markov chain Monte Carlo methods and variational methods.

\subsubsection{Markov-Chain Monte Carlo sampling}
\label{sec:orga88a69b}

In \textit{Markov chain Monte Carlo} (MCMC) sampling methods, the goal is to draw random samples from the normalized posterior distribution by using some possibly unnormalized representation of the posterior. Sampling methods do not attempt to approximate the posterior as any closed-form function. MCMC is the only practical method available that produces an unbiased estimate of complicated probabilistic models.

If it is run forever, an MCMC algorithm will produce an unbiased estimate of the posterior distribution if the distribution is well-behaved. Sometimes distributions have disconnected regions, and so the process fails to discover and sample from some regions. Additionally, MCMC algorithms take some time to explore the space - initial samples will be highly correlated with the initial sample and will not represent the distribution well. Because of this, the sampling process is usually allowed to run for a certain amount of time called a \textit{warm up period} before the samples are recorded. \\

\noindent\textit{Metropolis-Hastings.} Metropolis-Hastings (MH) algorithms are MCMC algorithms which have a particular method of sampling. The idea is to draw from a distribution called the \textit{proposal distribution}, and then to either accept or discard that draw according to a certain probability called the \textit{acceptance probability} \cite{MH}. The acceptance probability is a function of the unnormalized distribution that is carefully chosen to accept the appropriate distribution of draws according to the posterior distribution. Draws which move to higher density are more likely to be accepted. The proposal distribution at each step is a function of the previously accepted draw, making the sampling process into a Markov chain that explores the parameter space. There are many variants of MH, which differ mainly in their choice of proposal distribution.

In practice, MH with naive proposal distributions (such as Gaussian distributions centered at the last draw) can be very slow to explore high-dimensional spaces or spaces of difficult shapes, while more suffisticated variants like Hamiltonian Monte Carlo can achieve better performance. \\

\begin{figure}
	\centering
	\includegraphics[scale=0.23]{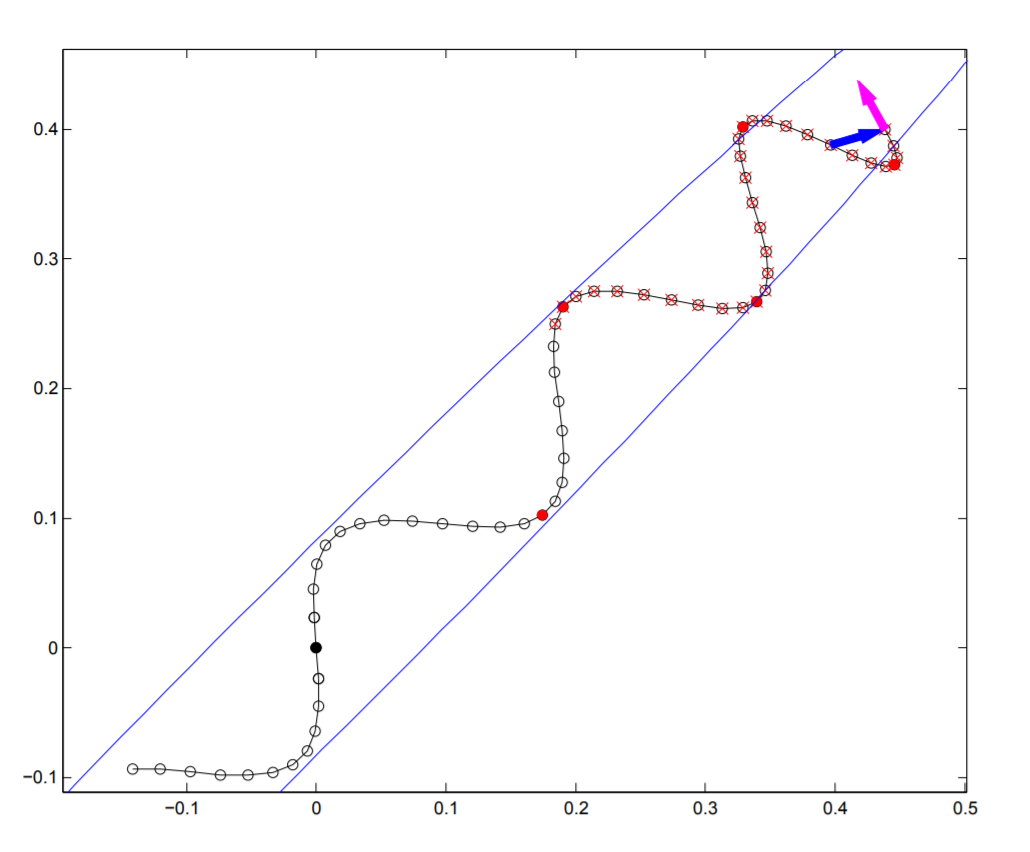}
	\caption{This shows the steps of a trajectory of a particle simulated by NUTS for an underlying posterior distribution whose contour is shown in blue. The axes represent two arbitrary dimensions of the same scale. The trajectory is terminated approximately when the particle would start moving back on itself. NUTS samples efficiently from the eccentric ellipse-shaped distribution. Figure taken from \cite{Hoffman2014TheNS}.}
	\label{fig:nuts-plot}
	\end{figure}
\begin{figure}
	\centering
	\includegraphics[scale=0.25]{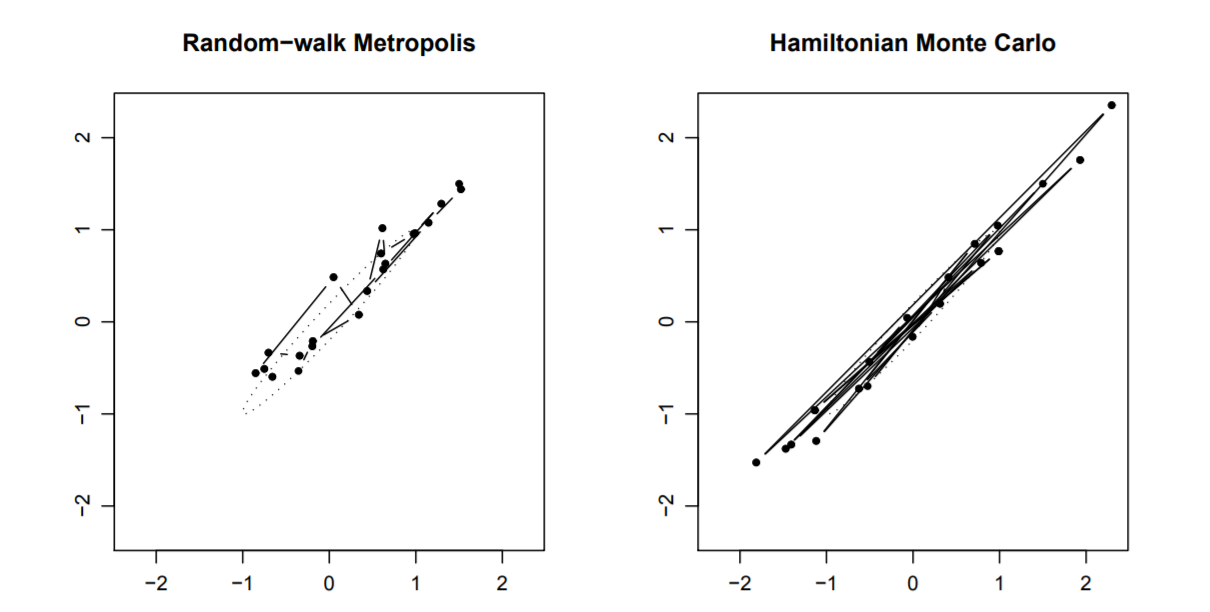}
	\caption{Shown here are the Markov chain trajectories for two different MCMC variants given the same joint distribution (whose elliptical contour is shown as dotted lines). The axes represent two arbitrary dimensions of the same scale. Successive samples are shown as dots connected by lines. The random-walk Metropolis algorithm has a Gaussian proposal distribution at each step, where the Hamiltonian Monte Carlo moves each step according to a particle simulated under Hamiltonian dynamics after an initial random momentum. The posterior is a difficult shape for the random-walk algorithm to traverse, but the gradient-informed HMC easily takes steps across the space. Figure taken from \cite{ermon_2019}.}
	\label{fig:hmc-vs-mcmc}
	\end{figure}

\noindent\textit{Hamiltonian Monte Carlo and the No-U-Turn Sampler.} Hamiltonian Monte Carlo (HMC) is a variant of MH that is particularly effective at quickly exploring difficult, high-dimensional distributions.

HMC works by using a connection between physics and probability theory:
If a particle with random momentum is moving around a space, a snapshot of the particle's location will be distributed according to a probability distribution that is related to the potential energy function of the space.
So, to draw a random sample from a probability distribution, we set up a space with the appropriate potential energy function, we simulate a particle moving and sample the particle's location.
This fits into the MCMC framework: the proposal distribution for the next sample is defined by the physical simulation from the last accepted sample.

The No-U-Turn Sampler (NUTS) (\autoref{fig:nuts-plot}) is an extension of HMC that automatically and dynamically selects runtime parameters of the HMC process, such as the length of each simulation between samples, to avoid inefficient particle movements.

HMC/NUTS is remarkably fast and accurate for high-dimensional problems, and is one of the standard workhorses in probabilistic programming. It tends to explore difficult and high-dimensional distributions more reliably than simpler variants of MH (\autoref{fig:hmc-vs-mcmc}).

MCMC algorithms are approximate and stochastic, and as a such it can be difficult to assess the correctness of their posterior estimates. There are a number of standard statistical tools for diagnosing problems, such as posterior predictive checks, which check the posterior samples against the available data \cite{ppc}, and Simulation-Based Calibration, which simulates data using the generative distribution and checks the quality of the inferred posterior distributions against the simulated parameters \cite{sbc}. There is also exist diagnostic metrics specifically for measuring the convergence of MCMC algorithms \cite{rank-normalization}.

In order to work, HMC/NUTS requires the gradient of the density function. This means that probabilistic programming languages which support HMC/NUTS usually implement automatic differentiation in their backend. It also means that those languages are restricted to differentiable density functions, which restricts the language semantics.

Other MCMC sampling algorithms require different structure from the density, such as the Hessian matrix at each point. Efficient Gibbs sampling requires a local statistical dependency set for each parameter called a \textit{Markov blanket}, which can either be gained from language design or by static or dynamic analysis \cite{Wingate2011NonstandardIO}.

\subsubsection{Variational methods}
\label{sec:org72d43e5}

\begin{figure}
	\centering
	\includegraphics[scale=0.3]{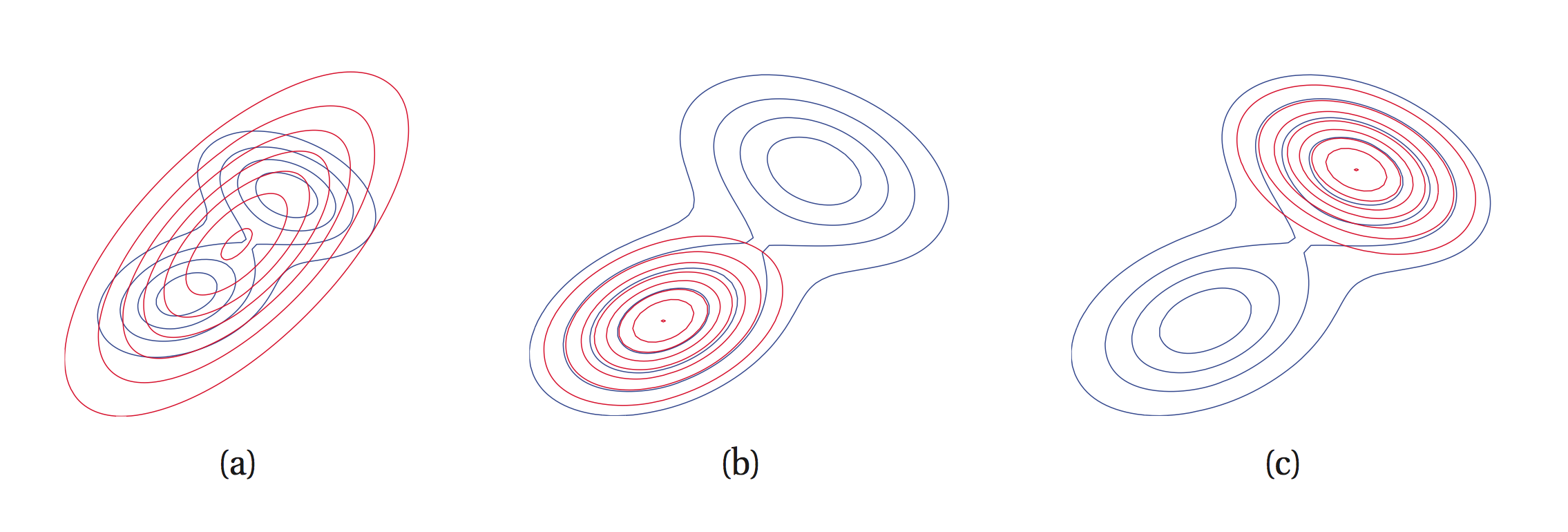}
	\caption{Shown here are the results of applying two variants of variational inference (red) to an underlying bimodal posterior distribution (black). (a) should VI with KL(P||Q), which tends to give an estimate more spread out than the underlying distribution. (b) and (c) show two possible result of VI with KL(Q||P), which tends to give more peaked estimates than the underlying distribution.}
	\label{fig:vi}
	\end{figure}

\noindent\textit{Variational inference} (VI) attempts to find an approximation to the posterior distribution within a restricted solution space of distributions called a \textit{variational family}. Parameters in the variational family are optimized to minimize a function called the Evidence Lower-Bound, which is a computationally tractable way of minimizing the KL-divergence, a measure of dissimilarity between the true posterior distributions and the approximation from the variational family. \autoref{fig:vi} shows example results of VI.

VI requires some variational family to be specified, and its accuracy can depend heavily on the choice of family. VI will not converge to the true posterior if it is not included in the family, and the resulting approximation will be biased depending on the exact definition of KL-divergence used. However, VI often converges faster than sampling methods, and sometimes scales better for large datasets in practice \cite{blei2017variational}.

Like MCMC algorithms, VI is an approximate method, and so assessing its correctness is also challenging. The standard posterior diagnostic tools such as posterior predictive checks and Simulation-Based Calibration also apply to VI. In addition, since VI is a (typically non-convex) optimization, converging to local optima is also a concern.

Some flavors of VI require first- or second- derivatives over the density, and so are subject to the same restrictions as MCMC algorithms which require the same.

\subsection{Static analysis}
\label{sec:org703ede1}

A wide variety of techniques are available for static analysis. Exploring this toolkit helps us understand the space of possibilities that can be applied in a probabilistic setting. This section briefly covers three topics that come up repeatedly in the current approaches for probabilistic programs, and which will likely be important for future advancements.
\subsubsection{Type systems}
\label{sec:org4813677}
\label{org9197d85}

\begin{figure}
	\centering
	\includegraphics[scale=0.32]{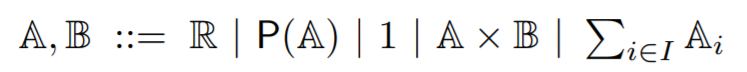}
	\caption{Shown are the types allowed by an example probabilistic programming language \cite{Commutative}. In English, it says: "Types named $A$ or $B$ will be one of: a real number $R$, a probability distribution over another type $A$, the unit value (which is only ever one value), a pair of two values of types $A$ and $B$, or a type that can be any of the types $A_i$ for $i \in I$." Effectively, this language allows real numbers, distributions, a singleton type, and any pair or disjunction of types. Figure taken from \cite{Commutative}.}
	\label{fig:commutative-types}
	\end{figure}
\begin{figure}
	\centering
	\includegraphics[scale=0.8]{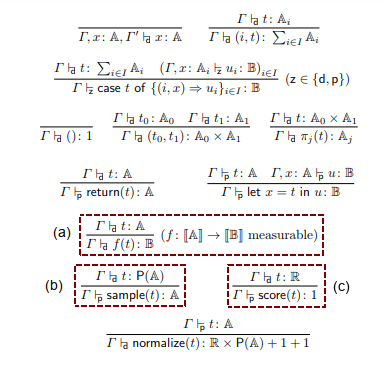}
	\caption{Shown are type rules sufficient to type-check the example probabilistic programming language \cite{Commutative}. A rule asserts that the statements over the line imply the statements below the line. $\Gamma$ represents the context available from the rest of the program. $\Gamma \vdash_x t : A$ means "the context implies that an expression t has type A". The subscript $x$ is used to distinguish between the implication of deterministic expressions with $\vdash_d$ and probabilistic expressions with $\vdash_p$. Non-probabilistic programs would only have $\vdash_d$, and would not include rules for $sample$, $score$ or $normalize$. \textbf{(a)} shows a non-probabilistic example, which intuitively reads: "For some expression $t$ with type $A$ from the context $\Gamma$, $f(t)$ is of type $B$ in the context $\Gamma$ for some measurable function $f : A -> B$." \textbf{(b)} shows the type rule for a $sample$ statement, which reads "For some expression $t$ which is a distribution over $A$, $sample(t)$ is a probabilistic expression of type $A$ in the context $\Gamma$. \textbf{(c)} shows the type rule for $score$, which reads "For some real number-valued expression $t$ in the context $\Gamma$, $score(t)$ is a probablilistic expression of which can only return one value (the unit value)." The other type rules define the rest of the type system in this way. The semantics of this language are discussed in a later section. Figure taken from \cite{Commutative}.}
	\label{fig:commutative-type-rules}
	\end{figure}

Types are descriptions associated with components of a program's syntax tree. An component's type describes properties of the value of the expression. For example, int and float are common names for types which correspond to integral and floating-point values.

Types are useful for restricting the set of programs that a compiler will allow. The compiler has a set of rules, called typing rules, which enumerate the ways that typed expressions and statements are allowed to fit together, and what their combined type is if they do fit together. If the compiler finds a point in a program which does not match any rule, the compiler rejects the program, and the program is said not to have type checked. For example, most compilers will reject statements like \(int\ x = 0.5;\) because they do not have a rule allowing the assignment of \(0.5\) to a variable of integer type.

\autoref{fig:commutative-types} shows the set of types allowed in the example probabilistic programming language from Staton \cite{Commutative}, and \autoref{fig:commutative-type-rules} shows the set of typing rules from the same language. These rules are sufficient to check if a program written in this language will type check and to infer the types of each part of the program.

If a program type checks, the programmer can be reasonably sure that the program is free of some class of bugs. Most compilers can promise that, for example, there are no floats masquerading as integers and each function is called with compatible parameters. The programmer can earn more peace of mind from more powerful type systems, such as ones that can express if a value falls within a certain range, or is a vector of a certain length. As type systems become more powerful, they can encode more of the specification of the program, and any program that successfully type checks is more likely to be correct. Encoding type systems so that they reject programs with important classes of bugs is a rich field of study.

If a program passes a type checker, the compiler has effectively proven something about the program. Writing type annotations into the program is akin to writing a proof about the program that can be checked automatically. Some compilers can infer some or all of the types of expressions based only on their usage - this is akin to the compiler automatically writing and checking a proof for the user. Compilers with advanced type systems often employ software such as \textit{Satisfiability Modulo Theory} (SMT) solvers as workhorses for automated theorem proving in this manner.

Types can also be useful tools for compilers to take into account in order to generate efficient code.

\subsubsection{Abstract Interpretation}
\label{sec:org6cee944}

\begin{figure}
	\centering
	\includegraphics[scale=0.38]{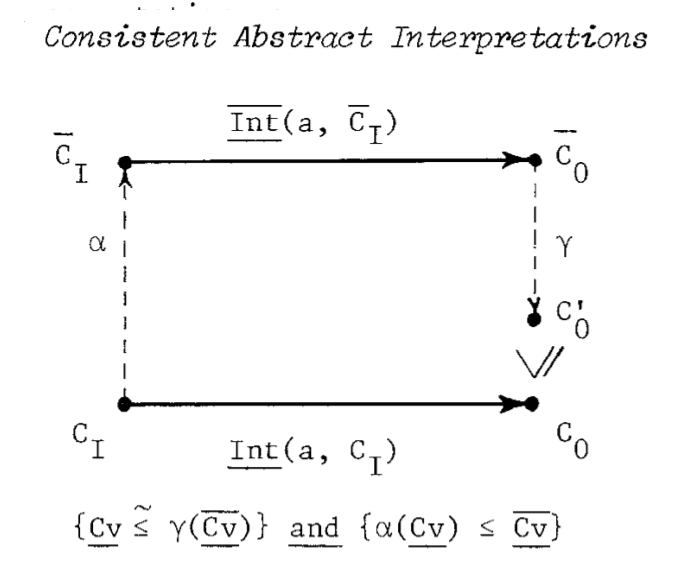}
	\caption{This is adapted from the original presentation of consistency from the 1977 Abstract Interpretation paper by Cousot \& Cousot \cite{AI}. $\alpha$ is the abstracting function, $\gamma$ is the concretization function, $Int$ is some program with maps inputs (subscripts $I$) to outputs (subscripts $O$). A line over a variable denotes that it is an abstracted version (the top part is abstracted). The $\geq$ sign denotes consistency, because there can be more than one concretization of an abstraction. An abstract interpretation is consistent if the result of flowing through the top, abstract path of the diagram produces a consistent result with flowing through the bottom, concrete part of the diagram. This is akin to a commutative diagram.}
	\label{fig:abst-int-consistency}
	\end{figure}

Abstract Interpretation (AI) is a foundational and general framework for reasoning about properties of programs and program parts. It provides a vocabulary and some useful mathematical tools. Some work has already been done to extend it into probabilistic settings.

In AI, the full meaning of a program is called the concrete semantics. The concrete semantics captures the whole behavior of a program when it is executed. The concrete domain is the space of possible concrete semantics for programs.

An abstract domain is a simplification of the concrete domain that only captures some interesting aspect of the computation. The function that maps from a concrete meaning to an abstract meaning is called an \textit{abstracting function}, and the reverse is called the concretization function. Given an abstracting function, a concrete program with concrete semantics can be transformed into an abstract program with abstract semantics. The abstract program does away with all details from the concrete program that do not affect the abstract domain. This abstract program is called an \textit{abstract interpretation} of the concrete program.

An abstraction interpretation is called \textit{consistent} if, for each possible input, the output of the abstract program is consistent with the output of the concrete program. A more formal statement is given in \autoref{fig:abst-int-consistency}.

For example (\cite{AI}), consider a program in an arithmetic language: \(-1515*17\), and suppose we care only about the sign of the result. We can use the abstraction of signs: the concrete domain of integers is abstracted into the abstract domain of signs. The abstract program is then \(-(+)*(+)\), where \(*\) is redefined to work on signs, and it is now easy to compute that the outcome will be \((-)\). This abstract interpretation is consistent since, for all such arithmetic programs, the abstract result will have the sign of concrete result.

This style of reasoning can be extended to reason about sophisticated properties of programs, and serves as the foundation for other, more specific approaches.
\subsubsection{Monotone Framework}
\label{sec:orgc728376}

\begin{figure}
	\centering
	\includegraphics[scale=0.45]{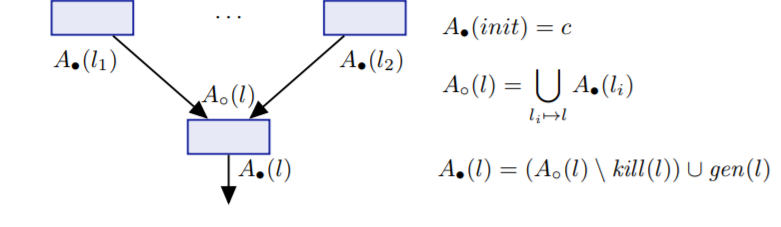}
	\caption{This shows how the Monotone Framework to join two branches of the control flow graph (such as after an $if-then-else$ statement) for some general abstract property. The transfer function is written in terms of $gen(l)$ and $kill(l)$, which denote the additions and removals of items from the abstract property set at the syntax element $l$. The $A_\bullet(l)$ shows the abstract property after executing the element $l$, and the $A_\circ(l)$ shows the abstract property before the execution of the element $l$ as the combination of the output $A_\bullet$s from previous nodes in the control flow graph. When this whole graph is drawn, the final $A_\bullet$ represents the final value of the abstract property for the whole program.}
	\label{fig:monotone-graph}
	\end{figure}
\begin{figure}
	\centering
	\includegraphics[scale=0.42]{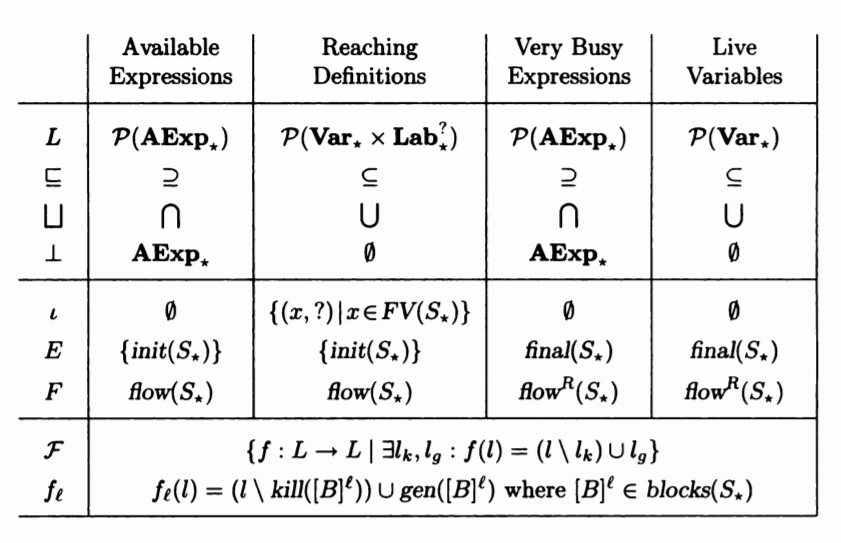}
	\caption{
Shown here are examples of the inputs that produce each of four common (and very useful) classical static analyses.
$L$ is the type of the property.
$\sqcup$ is the way that properties are rejoined after control-flow branches.
$\sqsubseteq$ and $\perp$ are the ordering relation and bottom element which define a lattice over properties, which is necessary for the Minimal Fixed Point Algorithm.
$f_l$ shows the transfer function in terms of $gen$ and $kill$ functions.}
	\label{fig:monotone-instances}
	\end{figure}

The Monotone Framework is a framework for applying the idea of abstract interpretation to programs in a mechanical way. Most of the standard set of program analyses fit nicely into this framework.

The Monotone Framework \cite{PrincipalsProgramAnalysis} is a process for finding abstract properties of the program at each point in its execution. The framework requires the following inputs:
\begin{itemize}
\item A function from a syntactic element (usually a statement) to a transformation of the abstract property. This is also called the \textit{transfer function}. The transfer function is usually described in terms of the \(gen\) function, which describes what is added to a property, and the \(kill\) function, which describes what is removed from a property.
\item A way of combining abstract properties from different program branches of control flow (such as an \(if\) statement's \(then\) and \(else\) branches). This is pictured in \autoref{fig:monotone-graph}.
\item The syntax tree of a program.
\item The control flow graph over the program, which specifies the branching structure of the statements.
\end{itemize}
The framework then produces a safe over- or under- approximation of the program property that holds at each point in the execution of the program, using the so-called Minimal Fixed Point Algorithm.

For example, consider the abstract property called the Reaching Definitions (RD) set. A reaching definition at a point is a definition (e.g. \(x = 5\)) which has not been overwritten up to this point in execution. The RD set is a set containing each reaching definition. For some programming language, we could describe how the syntax generates reaching definitions by changing a variable definition (the \(gen\) function) or removes reaching definitions by overwriting a variable definition (the \(kill\) function). We could define the way that the RD property is joined together across branches as taking a set union. We could then use the Monotone Framework to find the RD set at each point in a given program. The RD property is a prerequisite for building a dependence analysis graph.

\autoref{fig:monotone-instances} shows similar Monotone Framework inputs for three other classic analyses.

\subsection{Probabilistic Programming from a Static Analysis and Programming Languages perspective}
\label{sec:org35a7053}

In order to apply the rich theory of programming languages to probabilistic programs, we should first answer the fundamental question: what does a probabilistic program mean? When we type a probabilistic program into a text file, we know it represents an unnormalized probability distribution - but how exactly are the symbols that make up a program mapped to the mathematical object of a distribution? If we can answer this question with sufficient rigor, we will have a good foundation to apply existing static analysis tools to probabilistic programs.

A formal language semantics is a mapping from each valid program to its meaning. The space of all meanings is called the semantic domain.

This level of rigor allows us to reason with mathematical rigor about a number of questions about correctness in probabilistic programming:
\begin{itemize}
\item Language correctness: is the language specification complete and correct?
\item Correctness of posterior inference: will a posterior inference method given a program produce results consistent with the semantics of that program?
\item Program transformation: When a transformation (such as a compiler optimization) is applied to some program A resulting in program B, will the semantics of A match the semantics of B?
\item Program verification: What property can we prove about a program A that implies a property of the semantics of A?
\end{itemize}

For static analysis, we are especially interested in the help that formal semantics gives us with program transformation and verification. Keeping this formal foundation in mind helps to develop and prove correctness of static analyses.

There are two primary approaches\footnote{Axiomatic semantics would be a third example.} that have been used most often to build formal semantics for probabilistic programming languages: denotational semantics and operational semantics. The following sections introduce them and give an example language and semantics for each.

\subsubsection{Denotational Semantics}
\label{sec:org7a0e4ed}
\label{org77cda28}

A denotational semantics assigns a mathematical object to each component of the program, such that the meaning of the program is the composition of the meanings of its components. A denotational semantics is defined by writing \(\llbracket S \rrbracket = V\) for each element S of the syntax, where V is the semantic value of the element in the semantic domain. Denotational semantics is most nature to define for expressions as opposed to imperative statements, and works especially well for languages where everything is an expression with a value.
\begin{enumerate}
\item Example: A measure theoretic semantics for a sample/score language
\label{sec:org38e60f8}

\begin{figure}
	\centering
	\includegraphics[scale=0.26]{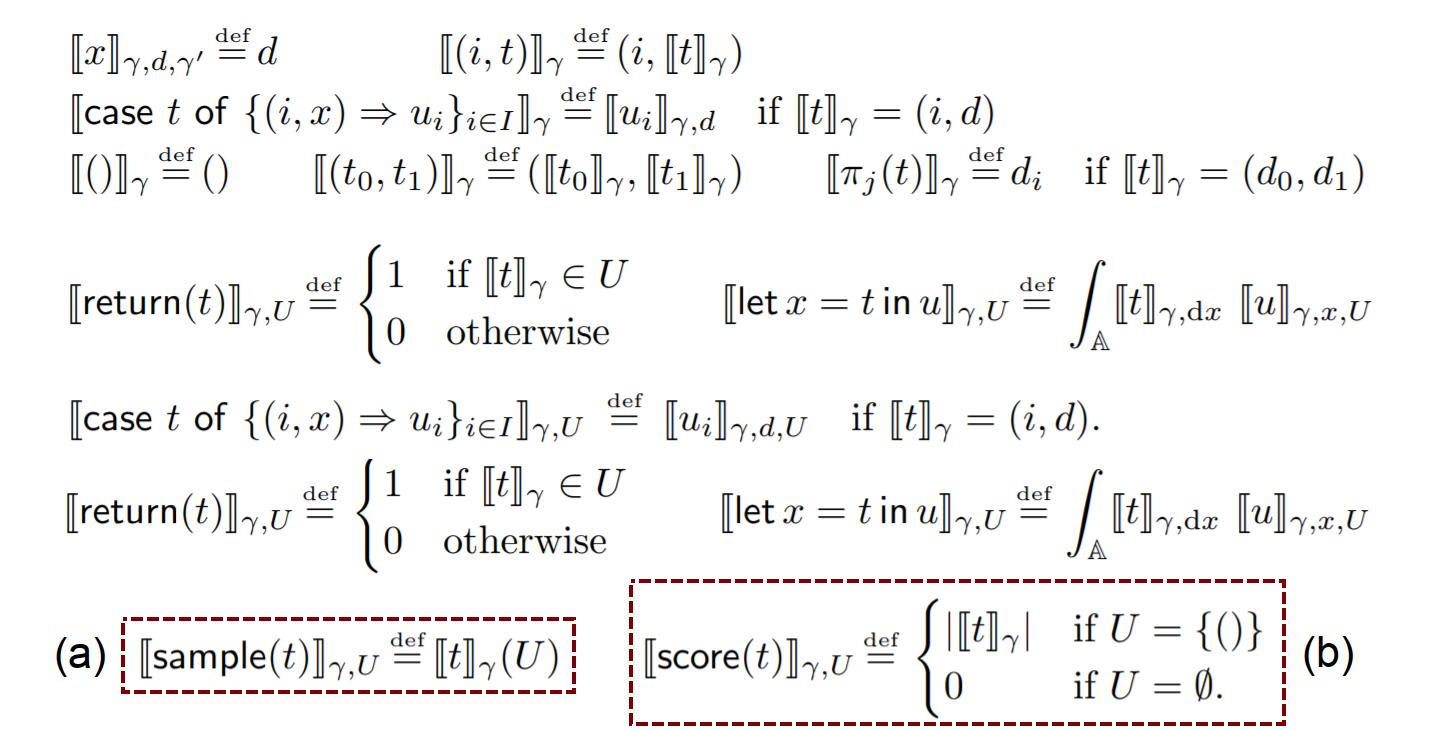}
	\caption{This is a standard way of representing the semantics of a language: each element of the syntax is mapped, by the semantic bracket function, to an object in the semantic domain on the right hand side. In this case, the semantic domain is a probability kernel $\llbracket \Gamma \rrbracket \rightsquigarrow \llbracket A \rrbracket = \llbracket \Gamma \rrbracket \bigtimes \Sigma_{\llbracket A \rrbracket} \rightarrow [0, \infty]$. Since the inputs (the program context $\gamma$ and the event $U$) are supplied as subscripts on the left hand side, the right hand side is in $[0, \infty]$. \textbf{(a)} shows the semantics of $sample$, and is read as "The density of $sample(t)$ with context $\gamma$ and event $U$ is the meaning of $t$ with context $\gamma$ evaluated at $U$." Intuitively, the likelihood of drawing $U$ from $sample(t)$ is $t(U)$, since $t$ must be a distribution (from the typing rule marked (b) in \autoref{fig:commutative-type-rules}). \textbf{(b)} shows the semantics of $score$, which is read as "The density of $score(t)$ with context $\gamma$ and event $U$ is the value of $t$ in the context $\gamma$ if there is an event $U$, otherwise $0$." Intuitively, $score(t)$ has density $t$ if it is evaluated to its return type $()$. The other statements define the rest of the semantics in a similar way.}
	\label{fig:commutative-semantics-boxed}
	\end{figure}

An example of an expression-only probabilistic programming language is given in \cite{Commutative}, whose type system is shown in \autoref{fig:commutative-types} in section \ref{org9197d85}. 

Staton \cite{Commutative} gives a denotational semantics for this language is given in terms of measure theory:
\begin{itemize}
\item A type \(A\) is interpreted in the semantics (written \(\llbracket A \rrbracket\)) as a \textit{measurable space}. A measurable space is a set along with a \(\sigma\)-algebra \(\Sigma_A\), which is a set of subsets of the space that can be interpreted as the set of subsets that can be formally assigned probabilities, called \textit{events}.
\item A deterministically-valued expression which depends on some surrounding context is a \textit{measurable function} from the context to the expression type. A measurable function is a well-behaved function such that the preimage of a measurable set is also measurable. This can be written: \(\llbracket \Gamma \vdash_d E \rrbracket : \llbracket \Gamma \rrbracket \rightarrow \llbracket A \llbracket\). The \textit{context} \(\Gamma\) is the set of variables bound in the surrounding program. \(A\) is the type of the expression \(E\).
\item A probabilistically-valued expression which depends on some surrounding context is a \textit{measurable kernel} from the context to the expression type. A measurable kernel is a mapping from the context to a measure on the output. This is written: \(\llbracket \Gamma \vdash_p E \llbracket : \llbracket \Gamma \rrbracket \rightsquigarrow \llbracket A \rrbracket\), which is equivalent to \(\llbracket \Gamma \rrbracket \bigtimes \Sigma_{\llbracket A \rrbracket} \rightarrow [0, \infty]\).
\item Since a whole program in this language is a probabilistic expression without a surrounding context, the whole program is a measure on the model variables.
\end{itemize}

The full denotational semantics for this language is shown in \autoref{fig:commutative-semantics-boxed}.

If we repeatedly apply these rules, we reduce a program from a syntactic tree to a mathematical object - in this case, a measure on the model variables.
\item Example: Probabilistic graphical models as a semantic domain
\label{sec:org6dcafe4}

Since Infer.NET programs correspond directly to directed, acyclic probabilistic graphical models, it would be reasonable to define a denotational semantics for Infer.NET with directed probabilistic graphical models as the semantic domain. Drawing a variable from a distribution would correspond to a node with edges to the nodes representing the distribution parameters. An example Infer.NET program is shown in \autoref{fig:Infernet-example}.
\end{enumerate}
\subsubsection{Operational Semantics}
\label{sec:orgc7fc604}

An operational semantics assigns each component of the program to a mathematical object, but also describes how the component operates on some global meaning of the program. Operational semantics tend to be convenient to define for statements which can perform side-effects that change the meaning of the program, rather than just representing a value.

\begin{enumerate}
\item Example: Trace sampling semantics for a probabilistic lambda calculus
\label{sec:orge031588}

\begin{figure}
	\centering
	\includegraphics[scale=0.35]{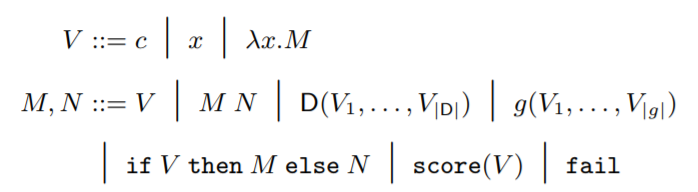}
	\caption{Shown here is the syntax of the untyped probabilistic lambda calculus in the same style as \autoref{fig:slicstan-stan-syntax}. The Score and Random features are akin to the \textit{score} and \textit{sample} functions from the denotational semantics example language in \autoref{fig:commutative-semantics-boxed}.}
	\label{fig:lambdacalc-syntax}
	\end{figure}

\begin{figure}
	\centering
	\includegraphics[scale=0.3]{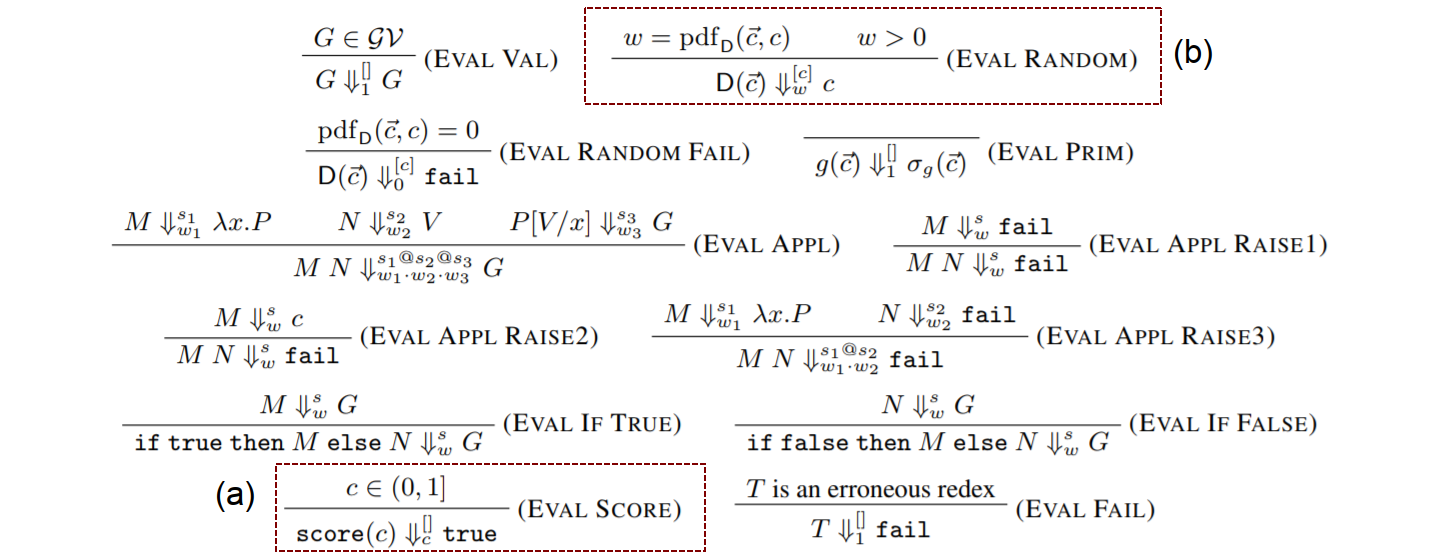}
	\caption{Shown here are rules for a call-by-value operational semantics for the untyped probabilistic lambda calculus. Rules are written in a similar style to typing rules, where the statements over the line imply the statement under the line. The $A \Downarrow^C_D B$ syntax is read as  "$A$ evaluates to $B$, adding $C$ to the record of draws with density $D$". \textbf{(a)} shows the rule for $score$, which reads "For some constant $c\in (0,1]$, $score(c)$ evaluates to $true$, does not add any draws to the record, and has density $c$." Score is used to explicitly scale the density at the present point in the distribution. \textbf{(b)} shows the rule for drawing random values from distributions, and reads "If $D(\vec{c})$ has density $w$ at $c$, $D(\vec{c})$ evaluates to $c$ with density $w$ and adds $c$ to the record of draws." The other rules define the rest of semantics in this way.}
	\label{fig:lambdacalc-operational-semantics}
	\end{figure}

Borgström \textit{et al.} \cite{LambdaCalc} define and provide an operational semantics for a probabilistic programming language called the untyped probabilistic lambda calculus. The syntax for the language is shown in \autoref{fig:lambdacalc-syntax}. Roughly speaking, a lambda calculus is a simple language centered around defining and applying functions. The language in \autoref{fig:lambdacalc-syntax} extends an untyped, call-by-value lambda calculus to include sampling random variables from distributions.

The full operational semantics for this language is shown in \autoref{fig:lambdacalc-operational-semantics}. The semantic domain is a pair of: (1) a record of all of the random samples drawn, and (2) the density of the unnormalized joint distribution evaluated at those samples points. The operational semantics defines how each element of the syntax contributes to this meaning by adding onto it. 

Borgström et al. go on to use this operational semantics to prove that the posterior samples from running the Trace-MCMC algorithm on a program is consistent with the semantics of the program. The ability to give such a formal correctness proof is an example of the advantages of formal semantics for probabilistic programming languages.
\end{enumerate}

\section{Properties of probabilistic programming languages salient for static analysis}
\label{sec:org3713d62}
Not all static analysis techniques make sense to apply to all probabilistic programming languages. This section enumerates a set of properties of probabilistic programming languages which, taken together, determine the applicability of a given static analysis approach. The author does not claim that this proposed set is complete or that it will suffice in the future, but at least that they are useful to keep in mind when exploring static analysis approaches. The properties of a small, diverse set of probabilistic programming languages are shown in \autoref{table:language-props}.

\subsection{Properties}
\label{sec:orgbc9affe}
\subsubsection{Discontinuous density functions}
\label{sec:orgb84490e}

Some languages and inference methods only work well (or at all) with distributions with densities that are continuous, differentiable, or differentiable to a higher order (as is the case with some MCMC variants). This requirement is sometimes incompatible with static analysis techniques that utilize discontinuous densities.

\begin{figure}
	\centering
	\includegraphics[scale=0.4]{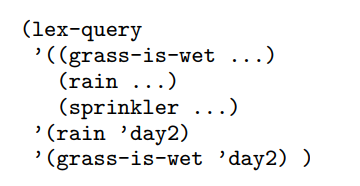}
	\caption{This is an example query from the Church language. The expression reads as: "Define a query over the model parameters \textit{grass-is-wet}, \textit{rain}, and \textit{sprinkler}, and return the distribution over \textit{rain} given \textit{day2} and conditioned on the predicate \textit{(grass-is-wet 'day2)}. Conditioning on this predicate almost always leads to a discontinuous density over the result of the query, even if the model parameters are continuous (they likely are not in this case). Figure taken from \cite{LambdaCalc}.}
	\label{fig:church-example}
\end{figure}

While it is possible to represent a Stan program with a discontinuous density function (for example, a program containing with control flow that depends on a model parameter), Stan does not claim to handle this case well with NUTS. The particle trajectory that HMC simulates to draw samples is difficult to simulate with a discontinuous energy function.
    The language Church, on the other hand, was built to support conditioning on predicates over parameters (\autoref{fig:church-example}), making discontinuities typical. Church generally uses non-HMC MCMC sampling.

Some static analysis techniques are not applicable to probabilistic programming languages that can not handle discontinuity, often because they utilize on parameter-dependent control flow that introduces discontinuity.

\subsubsection{Predicate query oriented}
\label{sec:orgbe84c08}

\begin{figure}
	\centering
	\includegraphics[scale=0.4]{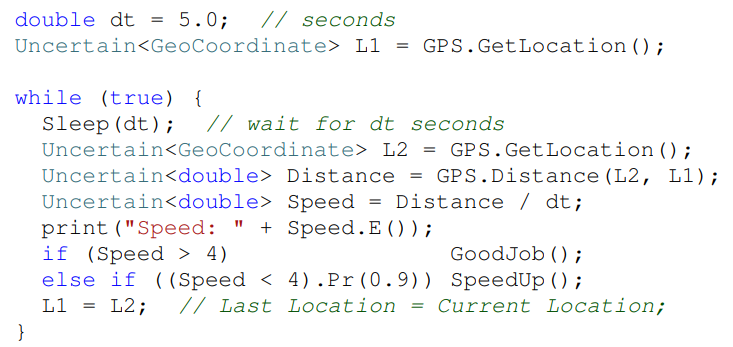}
	\caption{This shows an example usage of the C++ type Uncertain<T>. Distance and speed are defined to be random variables. They are used to determine which branch of the $if$ statement to follow, and so the only distributions that need to be infered are the distributions over the predicate queries ($Speed > 4$ and $(Speed < 4).Pr(0.9)$).  Figure taken from \cite{Bornholt2014UncertainAF}.}
	\label{fig:uncertainT-example}
\end{figure}

Some languages are designed to give a posterior probability of some predicate query over the parameters, rather than estimating the full posterior distribution of the parameters.
One example of such a language is Uncertain<T> (\autoref{fig:uncertainT-example}).

Some static analysis techniques rely on this predicate query to structure the computation, and usually involve reducing the program to focus on only answering the query. It is worth noting that languages which infer full distributions can benefit from these approaches in the context of verification, where a predicate query is specified or generated separately from the program definition.

\subsubsection{Sampling-based inference}
\label{sec:org14e6322}

Some languages specialize in sampling-based inference (MCMC), where other languages focus on other methods (such as variational methods).

\begin{figure}
	\centering
	\includegraphics[scale=0.45]{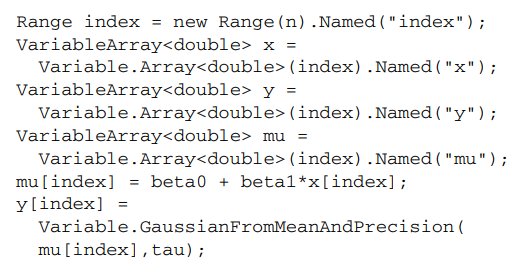}
	\caption{This shows an example Infer.NET model which defines a graphical model with variables $x$, $y$, and $mu$. Infer.NET's primary backend uses variational inference. Figure taken from \cite{InferNET}.}
	\label{fig:Infernet-example}
\end{figure}

Stan is an example of language focused on its MCMC backend, while Infer.NET (\autoref{fig:Infernet-example}) focuses on variational methods.

Some static analysis for probabilistic programming approaches specifically target the process of drawing samples in some way, and would not make sense to apply to languages focusing on other inference methods.

\subsubsection{Dynamic parameter declaration}
\label{sec:orgbe8ed1b}

\begin{figure}
	\centering
	\includegraphics[scale=0.45]{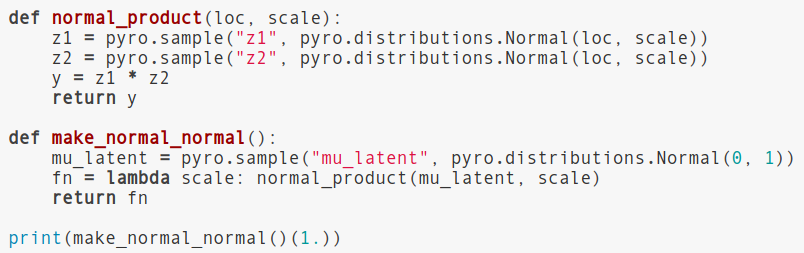}
	\caption{This shows a Pyro distribution. Pyro is written in Python for the Keras computational graph software platform. The function $normal\_product$ defines two new, locally-defined model parameters. Stan, for example, does not allow parameters to be declared dynamically inside functions. Figure taken from \cite{Bingham2019PyroDU}.}
	\label{fig:pyro-example}
\end{figure}

In most probabilistic programming languages, there is a constant set of model parameters defined over a program. This structure is not required in, for example, Pyro (\autoref{fig:pyro-example}). Pyro allows new model parameters to be declared depending on the branch of execution.

Some probabilistic programming-specific static analysis approaches lack the flexibility to be compatible with this level of flexibility, such as those which build and analyze graphical models to transform the program. These methods could potentially be extended to handle this flexibility by making conservative approximations.

\subsection{Properties of representative languages}
\label{sec:org6820134}

\autoref{table:language-props} shows examples of how the properties apply to real-world languages.

\begin{table}[]
\centering
\begin{tabular}{r|c|c|c|c|}
                                         & \shortstack{Discontinuous \\ density \\ functions} & \shortstack{Sampling-\\focused \\ inference} & \shortstack{Predicate\\ query \\ oriented}  & \shortstack{Conditional \\ variable \\ declaration} \\ \hline
Stan \cite{Carpenter2017StanAP}                                    &                  & \Checkmark  &            &                   \\ \hline
Uncertain\textless{}T\textgreater{} \cite{Bornholt2014UncertainAF} & \Checkmark       & \Checkmark  & \Checkmark &                   \\ \hline
Infer.NET \cite{InferNET}                                          &                  &             &            &                   \\ \hline
Church \cite{Goodman2008ChurchAL}                                  & \Checkmark       & \Checkmark  & \Checkmark & \Checkmark        \\ \hline
Pyro \cite{Bingham2019PyroDU}                                      & \Checkmark       &             &            & \Checkmark        \\ \hline
PSPs \cite{Kapoor2016ProbabilisticSP}                              & \Checkmark       &             & \Checkmark &        \\ \hline                         
\end{tabular}
\caption{This shows the properties for each of a small, diverse set of probabilistic programming languages.}
\label{table:language-props}
\end{table}

\section{Current techniques}
\label{sec:org1ab0f63}
\subsection{Overview}
\label{sec:org4c8b2e4}

There have been important and interesting contributions to the field of static analysis for probabilistic programming. They vary in their goal, approach and what they assume about the languages they target. It is not uncommon for a technique to be presented alongside an entirely new language on which to demonstrate it. This section attempts to categorize these contributions first according to their purpose and broad strategy, then to describe their approach and enumerate the language properties that they assume.

\subsection{Optimization}
\label{sec:org12d7932}

Perhaps the majority of current work focuses on the optimization of probabilistic programs.
\subsubsection{Avoiding work}
\label{sec:org28c9049}
\label{org08f168a}

The primary strategy for classical optimization is to avoid doing work during program execution in some way. This idea extends to probabilistic programming by avoiding work during posterior inference. The probabilistic semantics do require some additional considerations.

\begin{enumerate}
\item Program slicing and dead code elimination
\label{sec:org8640b39}

The idea of program slicing is to find the "slice" of a program which contributes to a value at some point. This task is highly related to the goal of dead code elimination, which aims to remove parts of the program which do not contribute to the final outcome of the execution. For our purposes, we can consider them to be the same problem, where dead code elimination preserves only the "slice" which contributes to the result of the program.

In the case of probabilistic programming, we seek to find the slice which contributes to the distribution that the program represents. We can apply the usual dead code elimination techniques to non-probabilistic portions of the program, but we can also sometimes eliminate probabilistic portions like extraneous model parameters. The examples below focus on eliminating unnecessary model parameters.\\

\noindent\textit{Example: Slicing probabilistic programs.} Program slicing is made a little bit more complex in the probabilistic setting because there can be indirect dependencies between model parameters which are not reflected directly by the syntax. Hur et al. \cite{Hur2014SlicingPP} provide a technique for program slicing which is aware of these indirect relationships, and can eliminate code to compute certain variables based on their independence from the model parameters being estimated. \\
\autoref{fig:slicing-before-and-after} shows how a program is transformed.
\begin{figure}
	\centering
	\includegraphics[scale=0.3]{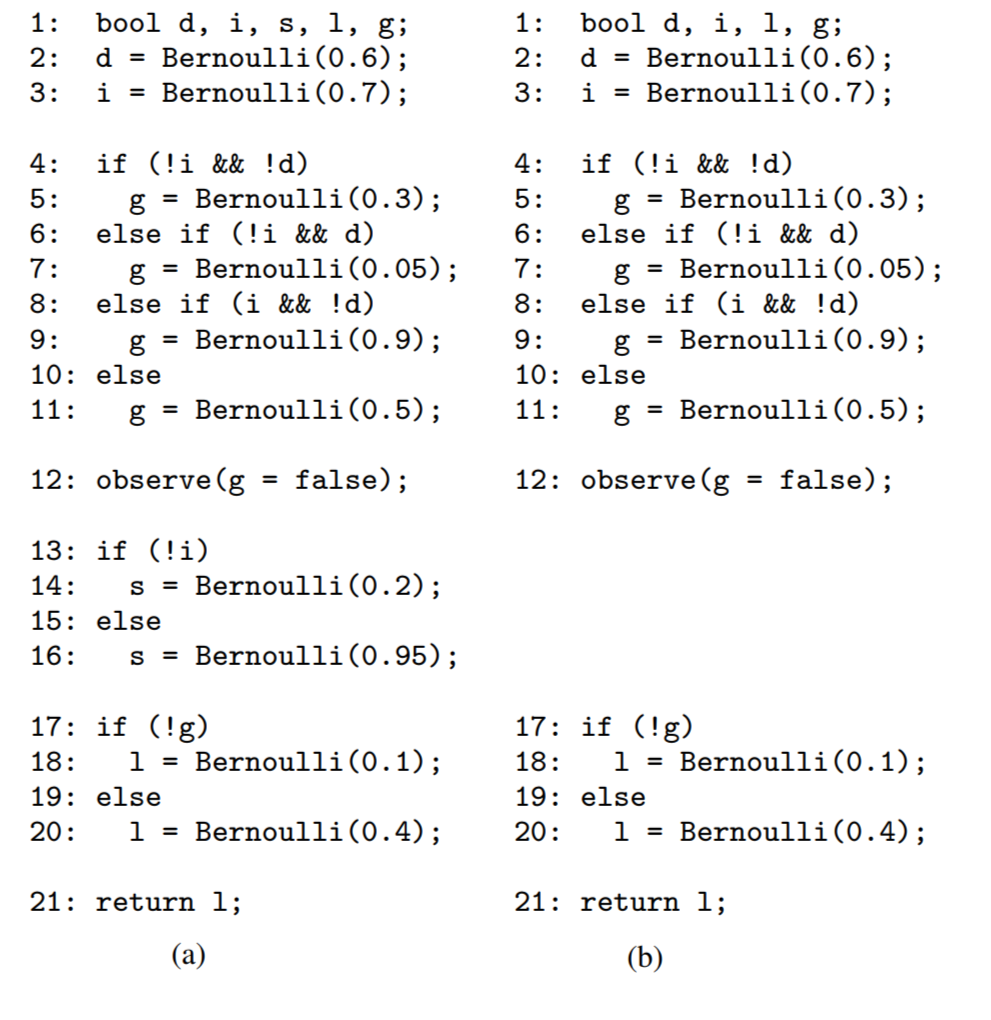}
	\caption{On the left is shown a probabilistic program which computes the distribution of a random variable $l$. Since only $l$ is returned, we can eliminate computations that assign to variables which do not effect $l$. \autoref{fig:slicing-graph} shows the probabilistic graphical model corresponding to the conditional dependence relationships between the variables in this program: using the usual rules for graphical models, we see that the value $s$ is independent of the value of $l$. The program on the right has the $s$ computations automatically removed but preserves the semantics of the full program.}
	\label{fig:slicing-before-and-after}
\end{figure}
\begin{figure}
	\centering
	\includegraphics[scale=0.33]{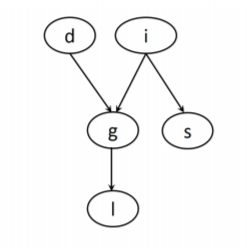}
	\caption{The directed graphical model corresponding to the program in \autoref{fig:slicing-before-and-after}. It encodes the following conditional independence: given $i$, $s$ is independent of $l$.}
	\label{fig:slicing-graph}
\end{figure}

\noindent\textit{Example: Hakaru.} Another example of program slicing is implemented into the compiler of the probabilistic programming language Hakaru. During Hakaru's "disintegration" phase, a program representing a joint measure is rewritten into a representation of a conditional distribution, and during this phase Hakaru drops pieces of the joint measure expression that do not contribute to the conditional distribution. \autoref{fig:hakaru-before} shows a Hakaru program before disintegration and \autoref{fig:hakaru-after-disintegration} shows the representation after disintegration.

This technique is most effective when the user specifies a query that does not require inferring the full posterior of all model parameters. Stan does not have a facility for the user to specify queries, and so must assume that all parameters are important. Church explicitly specifies its query, and so could easily take advantage of PPL-specific program slicing.

This technique requires a static analysis pass to find the dependency graph between variables. The dependency graph can be derived from the Reaching Definitions property found from the Monotone Framework. The dependency graph can be extracted from programs in any probabilistic programming language with varying levels of difficulty and specificity, and so program slicing can be applied in some form to any language.
\item Partial evaluation
\label{sec:org419e558}

     Rather than avoiding computation altogether, partial evaluation attempts to compute as much as possible at compile-time.
This is a very general and effective technique that is already applied as an optimization step in some probabilistic programming languages.

\begin{figure}
	\centering
	\includegraphics[scale=0.4]{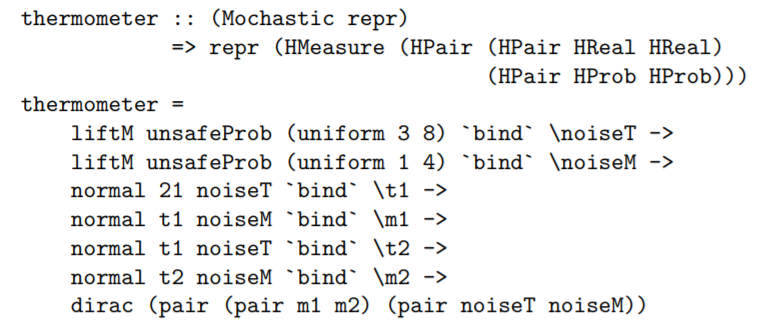}
	\caption{A Hakaru program. The first three lines are the type of the program. $noiseT$ and $noiseM$ are drawn from uniform distributions, $t1$, $t2$, $m1$ and $m2$ are drawn from normal distributions. The final line returns a joint distribution where $m1$ and $m2$ are observed variables and $noiseT$ and $noiseM$ are model parameters.}
	\label{fig:hakaru-before}
\end{figure}

\begin{figure}
	\centering
	\includegraphics[scale=0.4]{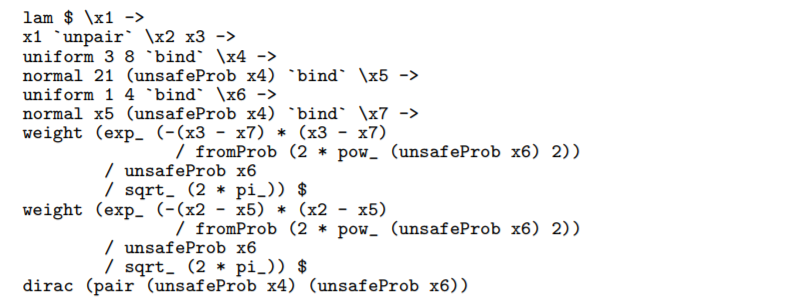}
	\caption{This is a Hakaru expression representing the program in \autoref{fig:hakaru-before} conditioned on $m1$ and $m2$ (named here $x2$ and $x3$). The program will also be sliced with respect to the output variables $t1$ and $t2$ (now named $x4$ and $x6$). This representation is confusing to read because it has been automatically generated.}
	\label{fig:hakaru-after-disintegration}
\end{figure}

\begin{figure}
	\centering
	\includegraphics[scale=0.4]{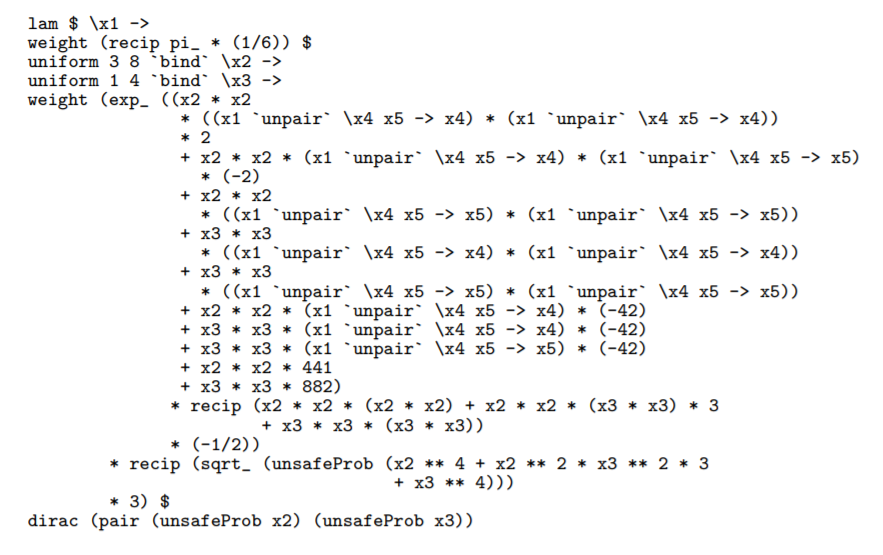}
	\caption{This is a Hakaru expression representing the conditional distribution in \autoref{fig:hakaru-after-disintegration} after being partially evaluated by Hakaru's "Simplification" transformation, which applies algebraic reductions with the Maple software package. Again, this representation is difficult to read because it is automatically generated.}
	\label{fig:hakaru-after-simplification}
\end{figure}

For example, the Hakaru language implements partial evaluation by translating pieces of the program into an algebraic language, passing it to an algebraic simplifier, and then tranlating back to the Hakaru representation. Hakaru calls this the \textit{simplification phase}. A Hakaru representation before simplication is shown in \autoref{fig:hakaru-after-disintegration} and after simplicification in \autoref{fig:hakaru-after-simplification} The representation of a Hakaru program before This affords Hakaru extra efficiency depending heavily on the choices of variable distributions and the strength of the solver.

The implementation of partial evaluation probabilistic languages is not significantly different than its implementation for non-probabilistic languages. The approach is to search for patterns of subexpressions in the program which can be reduced to some simpler form.

\item Simplification by abstract interpretation
\label{sec:org6182bc3}

When a probabilistic program is only being used to answer some query, it may be possible to simplify the program from the fully concrete domain into a more abstract domain that can still represent the query. This abstract program may allow for faster posterior inference \cite{Holtzen2018SoundAA}. This idea is covered in \label{orge5b6f0e} for the specific case of verifying a predicate query by abstracting the program into a predicate domain. In theory, it could be used to optimize programs for queries more complex than predicates but less than full parameter densities.
\end{enumerate}

\subsubsection{Inference-aware optimization}
\label{sec:org63532e3}

Another avenue for optimization, which is a opportunity unique to probabilistic programming, is to use program static analysis to improve the inference process. In some circumstances, probabilistic programs can be transformed to be more efficient for a particular posterior inference method. So far work on this approach is limited to improving the efficiency of sampling methods.\\

\begin{figure}
	\centering
	\includegraphics[scale=0.35]{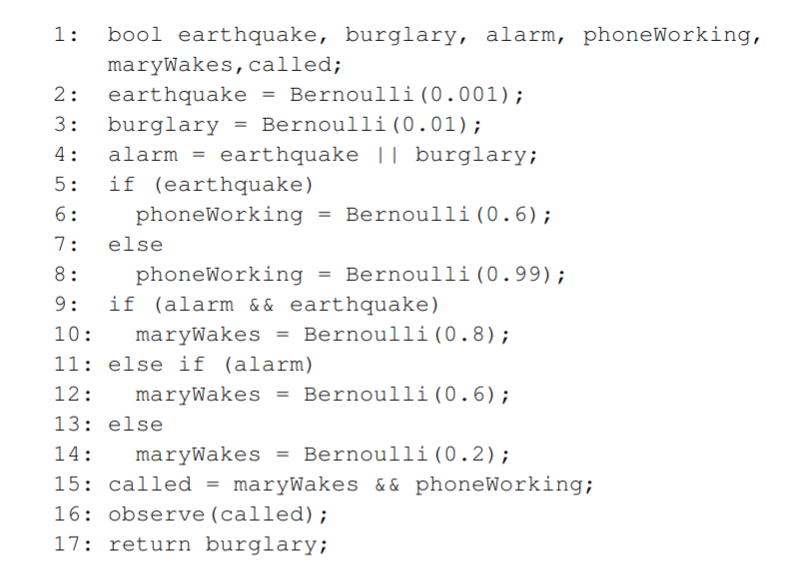}
	\caption{An R2 program before transformation. The sampling statements (e.g. line 6) are separated from the observation assertion (line 16), so some samples will disagree with the observations.}
	\label{fig:R2-before}
\end{figure}
\begin{figure}
	\centering
	\includegraphics[scale=0.35]{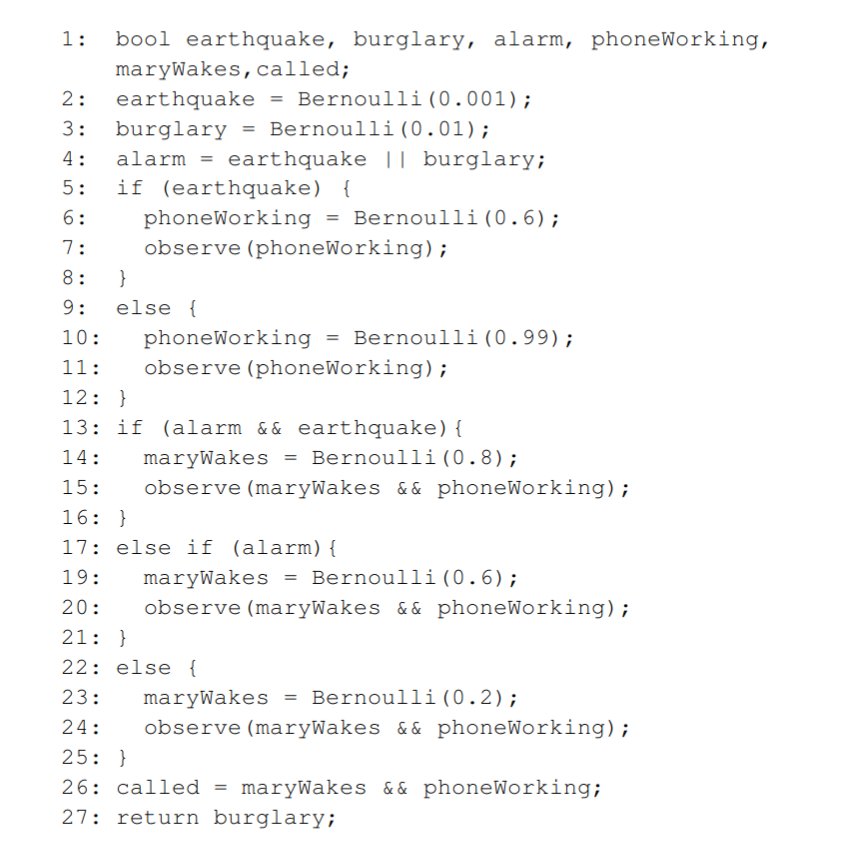}
	\caption{The R2 program after transformation. Each sampling statement is accompanied by the necessary conditions on the sample (for example, line 6 accompanied by line 7).}
	\label{fig:R2-after}
\end{figure}

\noindent\textit{Example: Optimizing sampling with R2.} Nori et al. \cite{Nori2014R2AE} introduced an inference algorithm called R2 whose main innovation is overcoming a particular inefficiency in drawing samples from a probabilistic program.

An example of the situation that R2 attempts to improve is shown in \autoref{fig:R2-before}. The example is written in a language specified alongside R2. In this program, some binary random variables are first sampled from distributions. Observations of those samples are then specified (line 16) in the form of a predicate assertion. A naive sampling process for this program would generate the samples from the Bernoulli distributions, but they would very often not satisfy the observed condition, and so they would be rejected and work would be wasted.

Instead of this naive sampling process, R2 first transforms the program to specify the constraints on the samples at same the time they are drawn, as shown in \autoref{fig:R2-after}. Then, instead of sampling from independent distributions and rejecting many of the samples, R2 directly samples from distributions conditioned on the constraints being true. Since the sampling process itself is aware of the restrictions on the samples, no samples are rejected.

The program transformation is computed by cascading the logical implications of the observations backward through the program by finding the weakest precondition necessary for each successive command. This is called the pre-image operator.

This technique relies heavily on restrictions in the demonstration language. The pre-image operator is not necessarily computable for general languages.
It is shown only for Bernoulli distributions, but the authors suggest that the approach could be extended to other distributions by truncating their support to match conditions on the samples.
This approach will probably only work with observations which reduce the support of the posterior.

\noindent\textit{Nonstandard interpretations.} Non-standard interpretations are alternative ways of reading a program which give a different semantics to code that is already written. A typical strategy for non-standard interpretation is to replace an existing data type with a new data type that contains additional information. A common example of this is automatic differentiation \cite{ad}.

Since non-standard interpretations don't necessarily happen at compile-time, they aren't strictly static analyses. However, since in the probabilistic programming context they may happen before the inference phase, they may effectively work at as static analyses, and so they are mentioned briefly here.

Non-standard interpretations have been used in the probabilistic program setting to add structure to the definition of the density function in order to enable an inference method. The most common example is automatic differentiation, which automatically provides a density gradient for inference methods such as HMC \cite{Wingate2011NonstandardIO}. Another example is \textit{provenance analysis}, which is a sort of ad-hoc dependence analysis, which can enable alternative MCMC algorithms \cite{Wingate2011NonstandardIO}. A third example is a system called AutoConj \cite{AutoConj}, which searches the code of a density definition for conjugacy structure to make inference methods like Gibbs sampling simpler and more efficient.

\subsubsection{Conclusions}
\label{sec:org6445b30}
Depending on the language, probably the most practical source of optimizations will be classical optimizations applied to the non-probabilistic programming specific portions of the pipeline, such as the density computation. Many classical optimizations can be applied without major modification.

The program slicing and partial evaluation ideas presented in this section are direct extensions of classical optimizations to probabilistic programming. Both of these ideas are likely to speed up any language they are applied to.

R2 represents interesting theoretical work on optimization that takes advantage of the actual structure of the probabilistic programming pipeline. However, it is unlikely to be practically useful until it is extended to be compatible with popular languages and inference techniques.

\subsection{Verification}
\label{sec:org9d4c374}
\label{org761247e}

In probabilistic programming, verification queries can be defined over probabilistic quantities. In such a case, we cannot necessarily answer the query by proof of disproof, but rather by estimating (or bounding) the probability that the query holds. One strategy for estimating the query probability is to simply sample from the program and observe how often the query is satisfied. This strategy can work, but the query may require rare events to be estimated, and so it may take an unreasonable number of samples to get sufficient certainty on the bound of the estimate. Most of the current work relating to the verification of properties of probabilistic programs deals with making this estimate feasible.

\begin{enumerate}
\item Dividing the space of executions
\label{sec:org1b537e6}

One strategy is to divide up the space of possible executions by the path taken through the program, and then evaluate each path separately. This strategy will also allow us to find formal bounds on the probabilities of validation predicates, given that we use a fairly restrictive language to make the proofs easier.

Consider the probabilistic program in \autoref{fig:paths-prog}. Our goal is to find the probabilities of the predicates on lines 11 and 12. Ideally, we could sample executions of the this program infinitely many times, taking new samples of the random variables each time, and we would know the predicate probabilities. Instead, the strategy will be:
\begin{enumerate}
\item Sample some executions of the program, keeping track of the paths of the executions. A path is defined by the control-flow decisions that are made throughout the execution, such as the branch taken at each \(if-then-else\) statement. Continue sampling these paths until we are confidence that, with high probability, the path of a newly sampled execution will fall into the already observed set.
\item Find a proven lower bound on the probability of a newly sampled execution having a path in the observed set of paths.
\item For each path in the observed set, compute upper and lower bounds on the probability of each validation predicate using the information that the path implies. For example, if the path includes the \(then\) branch of an \(if(X)\) statement, then \(X\) is known about all executions that take this path, and \(X\) may change the probabilities of the validation predicates. This extra information lets us compute tighter upper- and lower-bounds on the validation predicates.
\item We can now compute upper- and lower- bounds on the predicates over all paths by assuming the opposite extreme (certainly true or certainly false) is taken with whatever probability is not certain to be covered by the path set.
\end{enumerate}

This strategy relies heavily on the restrictions of the language used to demonstrate it. These restrictions include being limited to only basic arithmetic operations on variables, only integral and floating-point types, and limiting statements to assignments, conditionals and while-loops. The conditionals and validation predicates are restricted to linear assertions over reals and integers. Bounds on the probabilities of the predicates given paths are possible to compute because of these language restrictions. This strategy would be difficult to translate directly to more flexible languages.

The strategy also depends heavily on the use of control-flow with model parameter predicates to break up the execution space. This advantage would be lost in languages which do not support discontinuous density functions, and thus have trouble representing such predicates at all.

\begin{figure}
	\centering
	\includegraphics[scale=0.35]{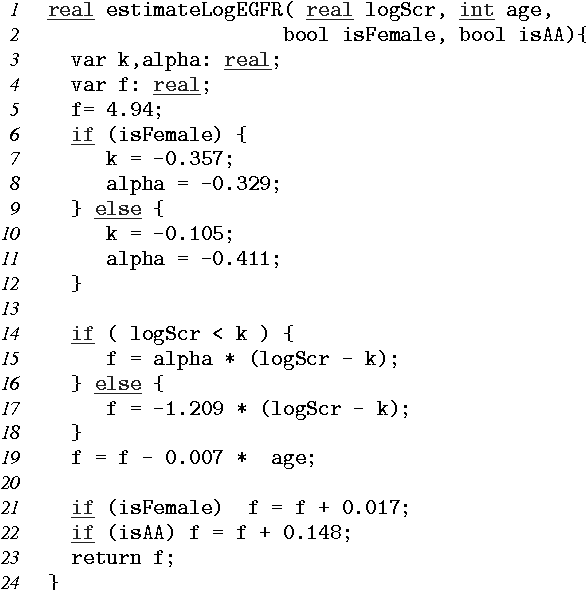}
	\caption{This is a deterministic part of the example program for the path sampling method.}
	\label{fig:path-prog}
	\end{figure}
\begin{figure}
	\centering
	\includegraphics[scale=0.35]{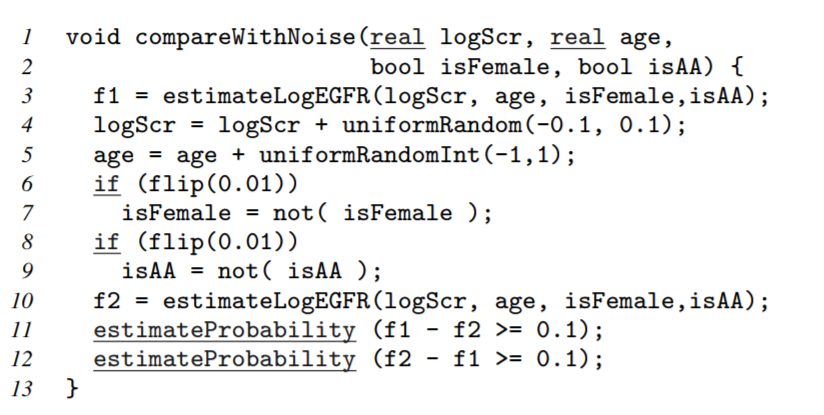}
	\caption{This shows a probabilistic program in the imperative language designed to demonstrate the path sampling method. The underlined statements define verification queries. The queries ask whether the sensativity of the $estimateLogEGFR$ function to its inputs meets a threshold. The definition of the $estimateLogEGFR$ function is shown in \autoref{fig:path-prog}. The key point from that function is that its behavior depends on branching conditions of its inputs.}
	\label{fig:paths-prog}
	\end{figure}
\item Probabilistic abstract interpretation
\label{sec:org2efc96d}

The idea of abstract interpretation can be extended to probabilistic programs. We can extend the original definition of abstraction consistency (shown in \autoref{fig:abst-int-consistency}) to \textit{distributional} consistency, which says that the the probability of an abstraction of the results of the concrete program should be identical to the probability of an abstraction resulting from the abstract program.

For the verification settings, we consider abstract interpretation of probabilistic programs into predicate domains. Predicate domains are semantic domains containing only boolean variables which correspond to the truth or falsity of predicates on variables in the concrete domain. The validation query can then be described in this domain. If we can construct an abstract program in the predicate domain which is consistent with the original program, we can perform posterior inference on the abstract program in order to perform the verification. 

\autoref{fig:sound-abs-before-and-after-only} shows an example of a probabilistic program before and after abstracting the program. The verification predicate we are considering is \(z = 0\), so this is included as a variable in the abstracted program. The abstract program encodes the observation that \(z\) is zero if either \(x\) is zero or \(y\) rounds down to zero. If the functions \(discrete\_dist\) or \(continuous\_dist\) are expensive to compute, operating on the abstracted program instead will be make this inference task feasible.

\begin{figure}
	\centering
	\includegraphics[scale=0.2]{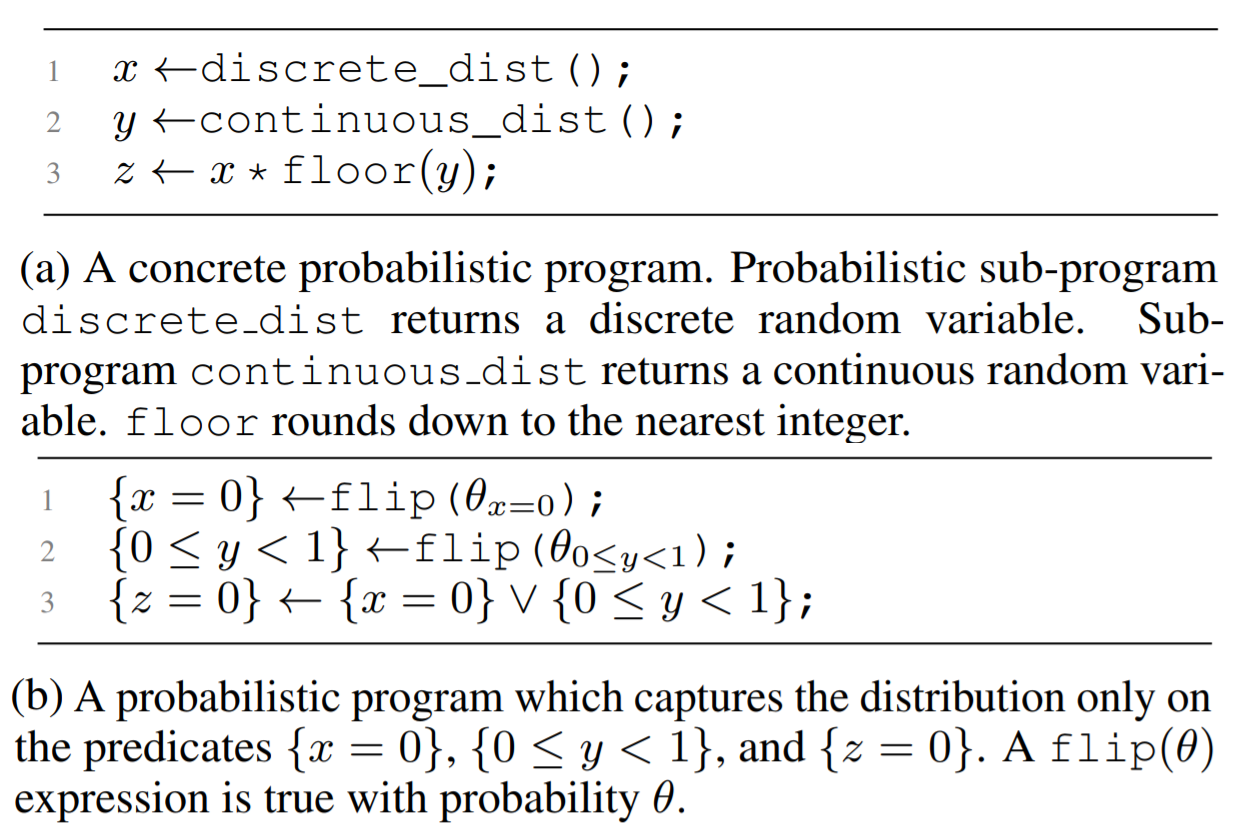}
	\caption{Example before (a) and after (b) abstracting a probabilistic program into the predicate domain, considering the query $z=0$. Figure taken from \cite{Holtzen2018SoundAA}.}
	\label{fig:sound-abs-before-and-after-only}
	\end{figure}
\end{enumerate}
\subsubsection{Conclusions}
\label{sec:org47fee98}

Verification for probabilistic programming is still quite a young idea, and is not yet especially practical - except in its classical form, which can still be used to prove non-probabilistic statements about non-probabilistic variables or probabilistic variables across a single evaluation of the density function.

Current probabilistic programming verification techniques mostly rely on restrictive demonstration languages for feasibility. Path sampling, for example, is only effective when the program includes many conditional branches which relate to the verification query, and can only handle very simple programming constructs. Probabilistic abstract interpretation has the potential to be a more general idea, but will likely require further development before it can be applied with any reasonable generality to real world probabilistic programming languages and scale to datasets.
\subsection{Usability}
\label{sec:org7488066}
\subsubsection{Compute away restrictions to the interface}
\label{sec:org6698df2}

Probabilistic programming languages tend to be much more user-friendly interfaces for machine learning than other approaches, since users can specify their models directly as programs. However, there are still major improvements to be made to the interface of the programs themselves. Some probabilistic programming languages have made design choices which favor a straightforward application their inference method of choice over ease of programming, such as Stan programs enforcing a block structure. Static analysis can sometimes offer the best of both worlds by providing an easier interface for users and translating it into the original programming language, retaining its ease of inference. This is the goal of usability methods. Another potential benefit of this translation step is to optimize the workflow that is automated away.

An example of this approach is SlicStan \cite{SlicStan}. SlicStan is a probabilistic programming language which is similar to Stan except that it relaxes some of the restrictiveness of Stan's syntax. SlicStan programs are compiled into Stan by re-organizing statements into a Stan program.

\begin{figure}
	\centering
	\includegraphics[scale=0.18]{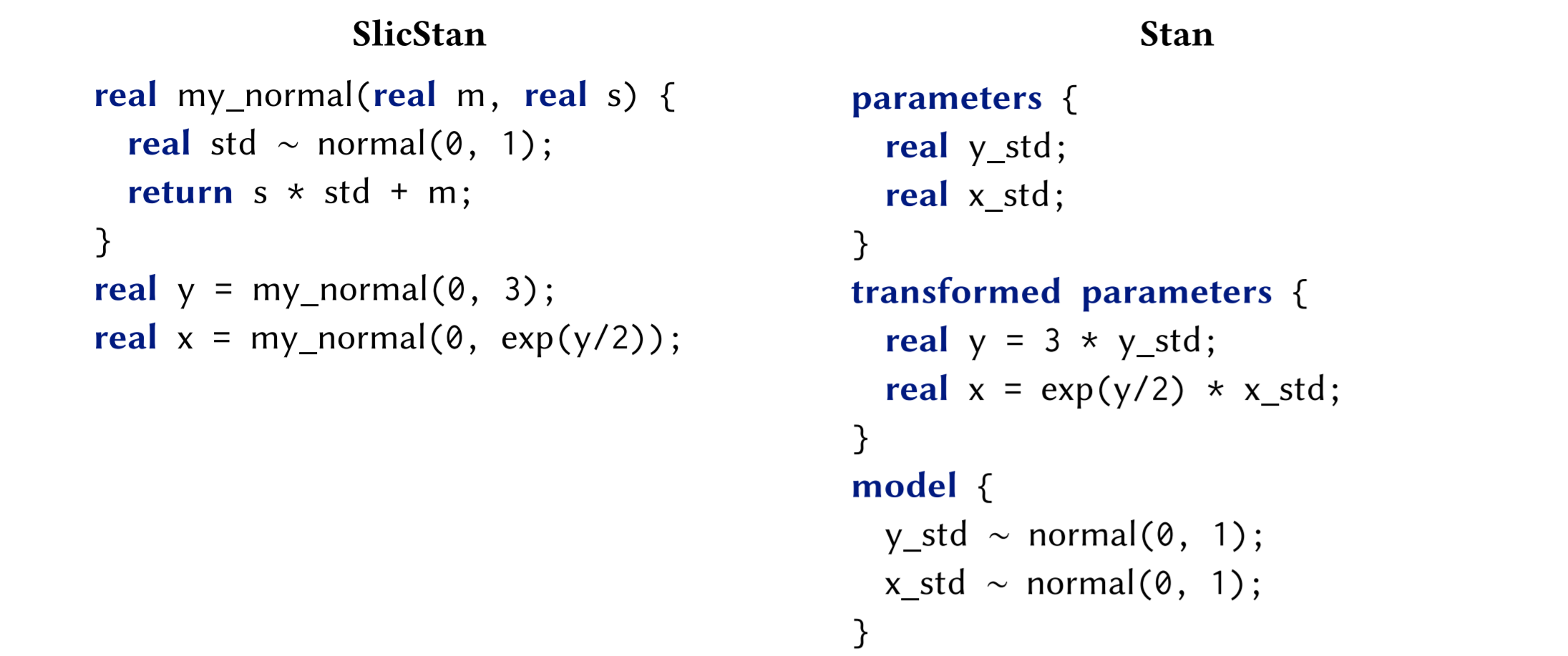}
	\caption{On the left is shown a SlicStan program, which allows users to forgo blocks and define more flexible functions. On the right is shown the equivalent Stan program that will result from compilation. Only the necessary statements are included in the $model$ block of the resulting Stan program.}
	\label{fig:slicstan-before-and-after}
\end{figure}

Stan programs are written in predefined blocks, such as the \textit{parameters} block for declaring model parameters and the \textit{model} block and \textit{transformed parameters} block for defining the log-density function. SlicStan's primary contribution is to not require these blocks. Instead, SlicStan allows users to mix statements that would otherwise be in different blocks, and sorts them into their optimal block when the program is compiled to Stan. \autoref{fig:slicstan-before-and-after} shows an example translation. By relaxing Stan's block system, SlicStan also extends Stan's user-defined function capabilities and makes composing programs more straightforward. One potential downside is that a user might lose some safety they would have by consciously structuring their program.

\begin{figure}
	\centering
	\includegraphics[scale=0.25]{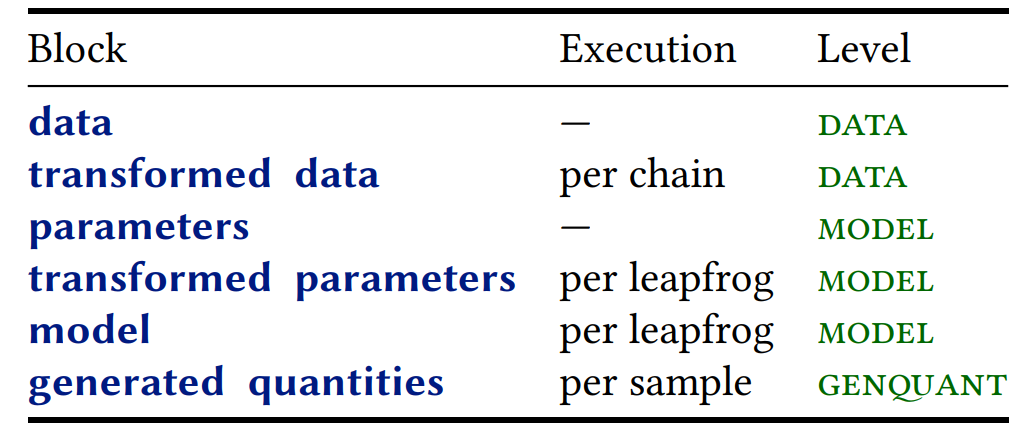}
	\caption{Each Stan block, along with the Stan phase during which it is executed and the level of SlicStan statement that will be assigned to it. "Per chain" means that the block is executed every time the NUTS algorithm is restarted on a new starting point, which typically occurs many times during posterior inference. "Per leapfrog" means that the block is executed every time the NUTS algorithm takes a step, which occurs many times for each chain. As a result, for example, it is less efficient to place a statement in the model block than in the transformed data block.}
	\label{fig:slicstan-blocks}
\end{figure}

During the translation, SlicStan's goal is to place each statement into the Stan block where that statement will be executed the fewest times while still executing correctly. \autoref{fig:slicstan-blocks} shows the execution time for each block. For example, since the \(model\) block will (typically) execute each time the NUTS algorithm takes a step, SlicStan only assigns statements to \(model\) that must be there, while any statement that can go into \(generated\ quantities\), which only runs once, should be placed there. 

\begin{figure}
	\centering
	\includegraphics[scale=0.2]{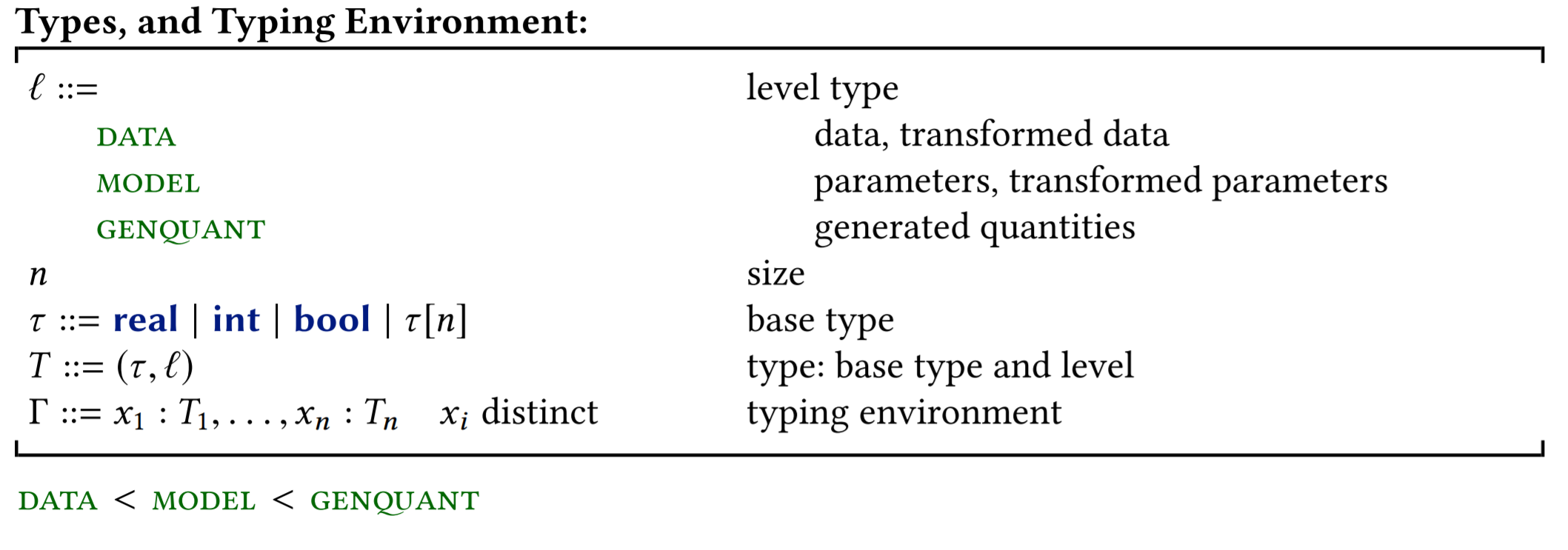}
	\caption{Enumerated here are the types of expressions in the SlicStan language. Each SlicStan expression is given a type $T$ which is a pair of a base type $\tau$ (taken directly from Stan) and a level type $\ell$. The level type indicates one of three options ($DATA$, $MODEL$ and $GENQUANT$), which indicate their block placement requirements.}
	\label{fig:slicstan-types}
\end{figure}

SlicStan works by augmenting Stan's type system for expressions with additional information (\autoref{fig:slicstan-types}).
SlicStan then infers the level type (also shown in \autoref{fig:slicstan-types}) for each statement using a set of typing rules.
Statements which must be in the model block, such as \$\(\sim\)\$-statements, are assigned the MODEL level type. Statements which receive information from MODEL level statements must be either MODEL or GENQUANT level statements, and statements which provide information to MODEL level statements must be either DATA or MODEL. Each statement is then assigned the least compute-intensive level that is permitted by these rules. This way, each statement is placed optimally without user input.

\subsubsection{Conclusions}
\label{sec:org6f11124}

SlicStan is the only major example of static analysis of probabilistic programs for usability, perhaps since user experience can also be improved by writing powerful libraries and documentation. However, presenting users with easier and more powerful interfaces has potential impact across all programming languages.
\section{Future directions}
\label{sec:org8867f7d}
\subsection{Extensions of current work}
\label{sec:org8542546}

This section enumerates some directions that would be natural extensions of the current work.   

\subsubsection{Optimization}
\label{sec:orgd5ff812}
\noindent\textit{Automatic transformation of models to use parallel computing hardware.} With sufficient understanding of the information flow within a probabilistic program, it should be possible to automatically transform probabilistic programs to take advantage of specialized computing hardware during posterior inference. This could include parallelism with the CPU, and GPU or TPU hardware. Some languages are already in a good position to implement this optimization, such as Edward and Pyro which target computation graph backends (Tensorflow and Keras,  respectively) that are already support specialized hardware. However, this optimization is applicable for most any language.

For example, if a static analysis method could show that the iterations of a \(for\) loop in a Stan \(model\) block were independent, the loop could be automatically parallelized. Parallelism within the computation of a density would be the easiest to achieve, but parallelism across other parts of the inference might also be possible for particular inference algorithms.\\

\noindent\textit{Passing additional structure to the inference algorithm.} Non-standard interpretation of source code can add to the semantics of a program, which can provide more information to an inference algorithm. Current examples include automatic differentiation providing gradients to HMC in Stan \cite{ad}; provenance analysis to enable alternative MCMC algorithms \cite{Wingate2011NonstandardIO}; and automatically discovered conjugacy structure to speed up various types of inference \cite{AutoConj}.

Advances in inference techniques could be coupled with new static analysis techniques which provide additional information from the program. For example, using dependence analysis on a Stan program to build a factor graph could enable Stan's backend to utilize the Sum-Product algorithm, which might enable Stan to automatically marginalize out discrete valued variables \cite{Bishop}. Dependence analysis could also be used to find the Markov blanket for each parameter, enabling efficient Gibbs sampling.\\

\noindent\textit{Partial evaluation of automatic differentiation.} Automatic differentiation is typically much more convenient and robust than user-provided derivatives, but it is sometimes less efficient. Attempting to find symbolic derivatives for expressions using algebraic solvers, in a similar style to Hakaru in \ref{org08f168a}, would sometimes speed up inference at the expense of compilation time and complexity.
\subsubsection{Verification}
\label{sec:org4d85e7f}
\noindent\textit{Relaxing the restrictions of verification methods.} Many current approaches to verification rely heavily on the restrictions of their demonstration languages. There are many potential ways to relax these restrictions which could allow these approaches to be applied more broadly.
     For example, in the paper introducing the R2 inference technique, Nori et al. suggest that R2's ideas could be applied to Stan - however, since R2 relies on observations in the form of predicate assertions rather than data points, R2's approach cannot be directly applied to Stan without expanding its notion of observation.\\

\noindent\textit{New probabilistic abstract interpretations.} Holtzen et al. demonstrated the use of predicate domains for verification (\ref{org761247e}) in a relatively simple environment. Their work could be extended by implementing more sophisticated versions of the pre-image operator to infer the weakest precondition of more types of predicates and statements. For instance, if an implementation of the pre-image operator were to include sufficient statistical knowledge, then perhaps statistical predicates could be included and reasoned about in the abstract domain.

\subsubsection{Usability}
\label{sec:org28a06dd}
\noindent\textit{Restrictive mode.} It could be useful to build restrictive subsets of existing probabilistic programming languages, which make it harder for users to make certain mistakes. This might be especially helpful to new users who might not understand the pitfalls of advanced features. This approach contrasts SlicStan, which aims to make the compilation process less restrictive.

For example, Stan includes a construct called \textit{reject()}, which eliminates a NUTS trajectory and effectively sets the density at that point to zero. This feature has the potential to introduce discontinuity into the density which NUTS cannot effectively estimate. A restrictive mode in Stan might disallow \textit{reject()} entirely.

In addition to being a helpful guide for usability, a restrictive mode might also be helpful for implementing many other static analyses discussed in this paper by disallowing problematic language features. For example, if some static analysis didn't work with recursive functions, a restrictive mode could disallow recursive functions.

\subsection{Unexplored directions}
\label{sec:org15a9cc2}

Listed here are directions that are of high potential and which are not directly reflected in the current literature. They are meant more to be discussion points than detailed plans.

\subsubsection{Statistically-aware user feedback}
\label{sec:orgacdf257}

There is potential for some system, supported by static analysis, to provide the user with feedback about the statistical model that their program represents. This could fall under the general category of usability, because statistical feedback may make probabilistic programming easier for users who are not absolute experts in statistics. It could also fall under verification, because the feedback may be phrased in terms of evaluating the truth of statistical properties on a program.

For example, the compiler could find the conditional independencies encoded in a program and provide a representation such as a graphical model to the user. The user could then check this model against their intent.

Other properties of a statistical model from the literature could also potentially be provided, a compiler could attempt to determine the statistical power of a model, letting a user understand if their dataset is likely to be sufficient to evaluate a query to a sufficient level of certainty.

In the extreme, statistically-aware user feedback could allow a probabilistic program compiler to take on the role of an interactive statistical assistant. An assistant may not be able to supply the main ideas of a model, but it could catch common errors and find mismatches between the program and the user's intentions. A conversation about the model could take place as an alternation of program iterations and compiler feedback.

\subsubsection{Validation of programs against other abstractions}
\label{sec:org038f111}

Typical program verification involves evaluating the truth of a predicate on the program variables. 

When the user writes a probabilistic program, they may have some other description of the model in mind, which are then made more concrete in the form of the program. For example, a user might be trying to encode a directed graphical model with some conditional independence structure, or they have some knowledge of the causal relationships between the variables. If the program is written correctly, this additional information will be encoded in the program.

These additional representations could be considered abstractions of the probabilistic program akin to the abstract programs in probabilistic abstract interpretation. They are not concrete enough to apply a posterior inference method, but they can be checked for consistency against a concrete program.

A user would provide an abstraction of their probabilistic model, such as a graphical model, in a specification language alongside their probabilistic program. A compiler could then check that the probabilistic program is consistent with the provided abstraction.

\subsubsection{Higher-level representations via transformation}
\label{sec:orga1cfd2e}

SlicStan provides a slightly higher level alternative interface for Stan that alleviates some of the programming complexities of writing Stan code.
It could be similarly useful to provide higher-level interfaces that alleviate some of the complex statistical considerations for the user.
Any piece of statistical knowledge that can be encapsulated into a higher-level interface is likely to be easier to write and more likely to be correct.

There could be domain-specific languages for a variety of fields which are specialized to easily and precisely express domain knowledge in that field. For example, a domain-specific language for computational genetics might include convenient manipulation of sequences-valued random variables, and its compiler might translate that representation into a Hidden Markov Model represented in Stan code.

Another example of could be automatically marginalizing discrete parameters.

\subsubsection{Data-scalable and data-absent analyses}
\label{sec:orgfd963dd}

In the current literature, static analyses generally assume that observations are integrated into the body of the program, and so this data is available to be part of the analysis at compile time. There are two potential issues with this assumption:
\begin{itemize}
\item Some analyses, especially verification methods, will not scale well when the data becomes very large.
\item In some probabilistic programming languages such as Stan, the dataset is not provided to the compiler along with the source code and is instead fed into the program after compilation.
\end{itemize}

Static analysis methods could be developed with special attention to these issues.
For example, interval analysis is a form of abstract interpretation where numeric types are abstracted as interval ranges representing their bounds. Interval analysis would be difficult to implement on a probabilistic program that has some values missing (in the case of absent data) or values that are resource intensive to traverse (in the case of large datasets). An alternative implementation of interval analysis could operate instead with some user-provided assumptions or an efficient process for summarizing the relevant features of dataset. Examples of these summaries might be dimensions, moments or bounds of a data variable.

\subsubsection{Posterior inference failure prediction and diagnosis}
\label{sec:org3124fa6}

One class of error discussed in section \ref{orgac79c64} is a failure of the posterior inference algorithm to converge to a reasonable approximation of the true posterior. Static analysis could do something to predict this type of error for a given program, perhaps also informed by the dataset. Any such analysis would be specific to the inference algorithm being run.

In general, implementation approaches could be:
\begin{itemize}
\item Searching for specific predictors of bad inference behavior. For example, if the posterior inference algorithm were HMC, the compiler could warn the user when the program is likely to have a discontinuous density or produce a non-Ergodic Markov chain.
\item Searching for sufficient indicators of good inference behavior (which will certainly be overly restrictive). For example, if the program is found to match a particular class of model known be supported by the inference engine, the compiler could be confident that the inference task for that program is viable.
\item Running posterior inference on automatically generated or subsampled datasets and checking for indications of failure. This may be a static analysis in the sense that it could be done before the real inference task, but dynamic in the sense that it runs the program with an inference engine.
\end{itemize}

\section{Conclusions}
\label{sec:org896ac90}

The current state of the field of static analysis for probabilistic programming is quite limited in scope. There is a disconnect between the language properties which theoretical static analyses assume and the languages which tend to scale well and have the most users. As a result, most of the immediate gains available to probabilistic programming language compilers are the same classical analyses available to other programming languages. However, current work has shown significant promise in the direction of probabilistic programming-specific static analyses.

Static analysis methods are already applied successfully in software engineering to provide high-level interfaces, automatic proofs-of-correctness, and optimizations without mental overhead. With advances in this field, applying static analysis methods to probabilistic programming has the unique potential to do the same for probabilistic modeling.

\section{Acknowledgements}
I would like to thank Professor Jeannette Wing, Professor Andrew Gelman, and Dr. Matthijs Vakar for their extensive feedback and guidance. I am also very grateful to Professor Itsik Pe'er for generously providing funding that enabled this work.

\nocite{*}
\bibliographystyle{plain}
\bibliography{paper}
\end{document}